\documentclass[fleqn,usenatbib]{mnras}

\usepackage{newtxtext,newtxmath}

\usepackage[T1]{fontenc}

\DeclareRobustCommand{\VAN}[3]{#2}
\let\VANthebibliography\thebibliography
\def\thebibliography{\DeclareRobustCommand{\VAN}[3]{##3}\VANthebibliography}

\usepackage{graphicx}	
\usepackage{amsmath}	
\usepackage{float}
\usepackage{array}
\usepackage{multicol}
\usepackage{multirow}
\usepackage{stfloats}
\usepackage{subcaption}

\title[Strategies for Accurate ePSF Modelling]{Strategies for Accurate Effective Point Spread Function (ePSF) Modelling on Undersampled Images}

\author[E. Godden and K. M. Blundell]{
Emma Godden$^{1}$\thanks{E-mail: emma.godden@physics.ox.ac.uk} and Katherine M. Blundell$^{1}$
\\
$^{1}$University of Oxford, Department of Physics, Keble Road, Oxford, OX1 3RH, UK
}

\date{Accepted for publication in RAS Techniques and Instruments.
}

\pubyear{\the\year{}}

\begin{document}
\label{firstpage}
\pagerange{\pageref{firstpage}--\pageref{lastpage}}
\maketitle

\begin{abstract}
Accurate modelling of the effective point spread function (ePSF) is essential for high-precision photometry and astrometry, particularly in undersampled imaging regimes. 
In this work, we build on a well-established ePSF modelling framework and its commonly used open-source Python implementation and demonstrate that several simple but effective modifications to existing ePSF modelling routines can significantly improve model accuracy.
We use synthetic ePSFs to generate simulated datasets of stellar images, allowing us to evaluate the accuracy of ePSF models and determine the scale of the pixel-phase errors in resulting flux and position measurements.
We systematically investigate how specific modelling choices affect ePSF accuracy, and evaluate the influence of oversampling, interpolation, gridpoint estimation, smoothing, star-sample distribution, and dithering on photometric precision. 
We apply our refined ePSF modelling routine to images from the Global Jet Watch observatories, demonstrating its improved ability to recover an accurate ePSF for real astronomical images. 
Our findings highlight the importance of tailoring the modelling approach to the specific characteristics of the instrument and detector, as well as to the nature of the available
imaging data used to construct the ePSF model.
These results provide practical guidance for optimising ePSF construction, thereby improving the reliability of photometric and astrometric measurements.
\end{abstract}





\section{Introduction}

Accurate measurements within observational astronomy rely on knowledge of the instrumental point spread function (PSF). For astronomical images, in order to measure the brightness of a star, or its exact position, we need to understand how the light from a point source is spread out and blurred by the instrument that produced the image. Different scientific goals make different demands on the images taken by an instrument. For single-object photometric analysis, it is advantageous for the light from the target object to be spread across a large number of pixels on the detector, giving us a very well-sampled PSF. For imaging surveys, the aim is to maximise the number of objects we can have in an image by having a large field of view and a compact PSF to avoid blending between objects. For some ground-based telescopes, it is often possible to still achieve a moderately sampled PSF in wide-field images, however for other ground-based and most space-based telescopes, wide-field images come with the challenge of having an undersampled PSF. 

There are significant challenges in obtaining accurate photometry and astrometry for images with an undersampled PSF. Aperture photometry methods are commonly used to measure the fluxes of point sources (stars) in wide-field images. These methods assume that it does not matter where exactly the light from a star falls, we only need to know within which pixels the light falls. Then we can estimate the flux of the star by simply taking the sum of the flux measurements in these pixels. The simplistic approach has some limitations. First, even for images with well-sampled PSFs, aperture photometry is unable to measure accurate fluxes for bright stars where the central pixels are saturated\footnote{This excludes detectors with a well-characterized saturation spill, where excess charge from saturated pixels is redistributed into neighbouring pixels in a predictable manner (as in the case of \textit{Kepler} \citep{Kepler_2020}). For such detectors, aperture photometry of bright saturated stars is still feasible, provided that all pixels containing the spilled flux are included in the aperture.}, or stars in crowded fields where individual pixels may contain flux from multiple stars. Second, complications in photometry and astrometry can arise due to the non-uniform nature of the sensitivity response across an individual pixel \citep{Lauer_1999}. Undersampled images can be particularly affected by this, causing a dependence on the flux reported by a pixel with where exactly within the pixel the peak of the PSF falls. This can lead to significant `pixel-phase' systematic errors in the aperture-measured flux and in centroid estimates that rely directly on the recorded pixel fluxes (i.e. flux-weighted, or centre-of-mass, centroids).

One possibility to improve the flux and position measurements in these cases is to use PSF-fitting photometry. PSF-fitting photometry is able to measure the flux and position of saturated stars by fitting the wings of the PSF and masking the saturated pixels. PSF-fitting photometry is also able to measure fluxes and positions of stars in crowded images by using iterative fitting procedures. In the case of undersampled images, provided that not all of the detectable light from a star falls within just one pixel, it is possible to obtain measurements of stars using PSF-fitting methods. A slight change in the centroid position of the PSF changes the relative flux that falls between adjacent pixels, so with knowledge of the PSF we can measure the position and the flux of the star. If the intra-pixel sensitivity variations are properly accounted for in PSF-fitting photometry, it is possible to minimise pixel-phase error systematics in the resulting measurements of stars even in very undersampled images. However, this is only possible with an accurate model of the instrumental PSF and the intra-pixel sensitivity variations.

A variety of PSF-fitting photometry methods have been developed and refined over the past few decades in an attempt to obtain accurate positions and fluxes of stars for a range of different instruments. Here, we focus our attention on the effective PSF (ePSF) photometry method developed by \cite{Anderson_King_2000}. 
The ePSF photometry method utilizes the fact that each image contains many samples of the PSF (i.e. via different stars in the field) that can be used to construct an empirical model of the PSF without any assumptions about its analytical form. \cite{Anderson_King_2000} make an important distinction between the instrumental PSF and the observed `effective PSF', the latter being the result of integrating the instrumental PSF over the individual detector pixels. This pixel-convolved version, the ePSF, is essential for accurate photometric and astrometric measurements. This version of the PSF is particularly useful for images with an undersampled PSF where pixel-phase systematic errors often appear in the lightcurves and positions of stars, as described in \cite{Lauer_1999}. Intra-pixel sensitivity variations are accounted for within the ePSF itself, hence avoiding pixel-phase bias systematics from occurring.

The purely empirical modelling of the ePSF makes this method adaptable to different instruments since no assumptions need to be made about the true shape of the ePSF, which is why we have chosen to focus on it here. It was initially developed for the WFPC2 camera of the HST (\cite{Anderson_King_2000}) and has since been used on a wide range of space-based telescopes (ACS/HRC: \cite{Anderson_King_2004}, ACS/WFC: \cite{Anderson_King_2006, Bellini_2018}, WFC3/UVIS: \cite{Anderson_2015, Anderson_Bedin_2017}, WFC3/IR: \cite{Anderson_2016}, Kepler/K2: \cite{Libralato_Feb2016, Libralato_Dec2016}, JWST NIRISS: \cite{Libralato_2023}, JWST MIRI \cite{Libralato_Mar2024}, JWST NIRCAM: \cite{Nardiello_2022}, Euclid: \cite{Libralato_Dec2024}, TESS: \cite{Nardiello_2019}, and CSST \citep{You_2025}). The ePSF method has also been adapted for use on ground-based telescopes (WFI@2.2m: \cite{Anderson_2006} and HAWK-I@VLT: \cite{Libralato_2014}). 

When being deployed for such a wide range of instruments, each new implementation of the ePSF routine requires careful adjustments to produce accurate results. This can be a challenging and time-consuming task for newcomers to the field who are wanting to use ePSF photometry for their observations. It is important to understand the accuracy and limitations of the ePSF modelling since using an inaccurate ePSF model can result in significant pixel-phase systematic errors in photometric and astrometric measurements. This work aims to elucidate generalised strategies for how to build an accurate ePSF model for any instrument, with a focus on undersampled images. 
In Section \ref{sec:ePSF} we give a summary of the ePSF method described in \cite{Anderson_King_2000}, and in Section \ref{sec:ePSF_code} we discuss its implementation in the \texttt{photutils.psf} subpackage of Astropy \citep{Bradley_2024}, highlighting differences and limitations relative to the original formulation. Section \ref{sec:simulations} describes how we use synthetic analytical ePSF models to generate simulated images to use in the ePSF modelling tests. Section \ref{sec:ePSF_building} investigates how different parts of the ePSF routine impact the accuracy of the ePSF model and resulting photometric and astrometric measurements. We discuss how some simple adaptations to the modelling routines discussed in \cite{Anderson_King_2000} and implemented in \texttt{photutils.psf} can significantly improve the accuracy of the ePSF modelling. Finally, in Section \ref{sec:GJW}, we demonstrate the application of the improved ePSF modelling routine to images from the Global Jet Watch observatories.

\section{A Reminder on the Effective PSF}
\label{sec:ePSF}

This section aims to summarise the ePSF formalism and modelling routine as described in \cite{Anderson_King_2000}, however the reader is recommended to refer back to this work for further details.

\subsection{ePSF Definition}
\label{subsec:epsf_def}

Each star in an astronomical image is a sample of the instrumental PSF scaled by the star's flux. However, the image of the star is not this instrumental PSF directly, but a collection of discrete pixels where the flux value of each individual pixel is the integral over the portion of the instrumental PSF which overlaps the pixel combined with the intra-pixel sensitivity profile. The intra-pixel sensitivity profile accounts for the fact that the response of a pixel is not uniform and there is a dependence on where precisely within a pixel a photon hits \citep{Lauer_1999, Mahato_2018}. We can write down the flux value of a pixel centred at $(i, j)$ in the vicinity of a star centred at $(x_{\ast}, y_{\ast})$ with flux $f_{\ast}$ as:

\begin{equation}\label{eqn:pix_flux}
    P_{ij} = f_{\ast} \int_{-\infty}^{\infty} \int_{-\infty}^{\infty} R(x-i, y-j) \Psi_I (x-x_{\ast}, y-y_{\ast}) \mathrm{d}x \mathrm{d}y + s_{ij}
\end{equation}

\noindent where $R$ is the intra-pixel sensitivity profile, $\Psi_I$ is the instrumental PSF, and $s_{ij}$ is the background value in that pixel.

The most important lesson from \cite{Anderson_King_2000} is that it is not necessary to try to deconvolve the instrumental PSF from the intra-pixel sensitivity profile. The instrumental PSF is never directly observed, so photometric measurements of stars in an image with the instrumental PSF would require re-doing the integration undone by the deconvolution. Instead, we should consider the `effective PSF' (ePSF). We can re-write Equation \ref{eqn:pix_flux} as:

\begin{equation}
    P_{ij} = f_* \int_{-\infty}^{\infty} \int_{-\infty}^{\infty} R(x, y) \Psi_I (x- \Delta x, y- \Delta y) \mathrm{d}x \mathrm{d}y + s_{ij}
\end{equation}

\noindent where we have changed coordinates to be in the reference frame of the pixel, $(x, y) \rightarrow (x-i, y-i)$, and defined the displacement of the pixel from the centroid of the star as $(\Delta x, \Delta y) = (x_* - i, y_* - j)$. We can then define the ePSF as a function of $(\Delta x, \Delta y)$:

\begin{equation}\label{eqn:epsf}
    \Psi_E (\Delta x, \Delta y) = \int_{-\infty}^{\infty} \int_{-\infty}^{\infty} R(x, y) \Psi_I (x - \Delta x, y - \Delta y) \mathrm{d}x \mathrm{d}y .
\end{equation}

Using the ePSF, the flux value of a pixel in the vicinity of a star simply becomes:

\begin{equation}\label{eqn:pix_flux_epsf}
    P_{ij} = f_{\ast} \Psi_E(\Delta x, \Delta y) + s_{ij} .
\end{equation}

Therefore, if we have a model for the ePSF we can easily fit the pixels in an image of a star to provide a measurement of the flux and centroid position of the star. In comparison to other PSF photometry approaches which model the instrumental PSF directly, no time-consuming integration of the ePSF model is required in this fitting process. We simply look up the ePSF values at different $(\Delta x, \Delta y)$ to find the best fit parameters.

\subsection{ePSF Modelling}
\label{subsec:epsf_modelling}

To perform ePSF fitting photometry, we first need an accurate model of the ePSF. Fortunately, each pixel within each star on the image is a sample of the ePSF at a specific $(\Delta x, \Delta y)$. Re-writing Equation \ref{eqn:pix_flux_epsf}, the ePSF for a given pixel at its corresponding $(\Delta x, \Delta y)$ is

\begin{equation}\label{eqn:epsf_pix_flux}
    \Psi_E(\Delta x, \Delta y) = \frac{P_{ij} - s_{ij}}{f_*} .
\end{equation}

If we have a set of stars with known fluxes and centroid positions, their pixel values can be used as samples to construct the ePSF. However, measuring stellar fluxes and positions requires an ePSF model in the first place. In order to accurately model the ePSF we need accurate measurements of stars, but accurate measurements of stars require an accurate model of the ePSF. \cite{Anderson_King_2000} resolved this degeneracy by iterating between measuring the ePSF and measuring the stars. Each new measurement of the ePSF improves the measurement of the stars and visa versa. They take an empirical approach in which the ePSF is represented as a table of values over a uniform grid of $(\Delta x, \Delta y)$ points, referred to as gridpoints. The ePSF at arbitrary $(\Delta x, \Delta y)$, which may be in between gridpoints, is found by interpolation between the tabulated values. The main steps used to generate this empirical ePSF model are summarised below.


\renewcommand{\labelenumi}{\arabic{enumi}.}
\renewcommand{\theenumi}{\arabic{enumi}}

\begin{enumerate}
    \item From a set of dithered images, select a sample of isolated stars with high signal-to-noise ratios. Make initial estimates of the fluxes and positions using alternative methods (e.g. centre-of-mass centroid and aperture photometry).
    \item Construct a table of ePSF values defined on a uniform grid of $(\Delta x, \Delta y)$ positions. Initially, all ePSF values should be set to zero.
    \item \label{item:pixels_to_ePSF_samples} For each sample star, convert each pixel value into a sample of the ePSF at its corresponding $(\Delta x, \Delta y)$ position using Equation \ref{eqn:epsf_pix_flux}. 
    \item \label{item:sample_residual_calc} For each ePSF sample, calculate the difference between the measured sample and the current ePSF model. In the initial iteration, the ePSF model is null so the residuals are simply the sample values. 
    \item For each gridpoint, take all the residual samples within a 0.25-pixel square and calculate the average residual. Adjust the corresponding ePSF table value by this average residual.
    \item The assumption is made that the ePSF model must be smooth. Enforce this by convolving the tabulated ePSF values with a pre-defined smoothing kernel. 
    \item \label{item:re-centring} Ensure that the centre of the ePSF lies at the centre of the empirical model. The definition of the centre of the ePSF is somewhat arbitrary, but must be kept consistent.
    \item Iterate through steps \ref{item:sample_residual_calc} to \ref{item:re-centring} several times. Five iterations tend to be sufficient.
    \item \label{item:remeasure_stars} Use the improved ePSF model to re-measure the fluxes and centroids of stars. Evaluate the pixel-phase error in the flux and centroid measurements, if it is sufficiently small then the ePSF model is complete. If the pixel-phase error is still significant, continue with steps \ref{item:calculate_averages} \& \ref{item:iteration}.
    \item \label{item:calculate_averages} Calculate the averaged fluxes and positions of the same stars observed in the dithered images to be used as un-biased measurements. The averaged positions must be calculated in a common reference frame and transformed back to the individual frame coordinates.
    \item \label{item:iteration} Repeat steps \ref{item:pixels_to_ePSF_samples} to \ref{item:remeasure_stars}.
    
\end{enumerate}

\section{ePSF Modelling Code Implementation}
\label{sec:ePSF_code}

An open-source version of the ePSF modelling and fitting routine is currently maintained in the \texttt{photutils.psf} subpackage of Astropy in the \texttt{EPSFBuilder} and \texttt{EPSFFitter} classes \citep{Bradley_2024}. This implementation has some differences in comparison to the original routine described in \cite{Anderson_King_2000}. These differences and their implications for ePSF modelling and fitting accuracy are summarised below.

\paragraph*{Iterations.}
In \citet{Anderson_King_2000} there are two levels of iteration in the ePSF-building routine. There is an overall iteration between measuring the ePSF and measuring the fluxes and positions of stars. Within each ePSF measurement there is an additional iteration between the adjustment of ePSF values based on the residuals, and the smoothing, re-centring, and normalisation of the ePSF. This internal iteration step only iterates over the re-centring step in \texttt{EPSFBuilder}. However, this difference is not expected to affect the final ePSF model, since both approaches iteratively refine the ePSF and stellar parameters toward the same self-consistent solution. The \texttt{EPSFBuilder} routine may simply require a larger number of outer iterations for the ePSF model to converge to an equivalent result.

\paragraph*{Re-centring.} 
In \citet{Anderson_King_2000}, the ePSF is defined to be symmetric about its centre, and any offset is corrected by shifting the sampling grid so that the ePSF values at $\pm 0.5$~pixel along each axis become equal. In \texttt{EPSFBuilder}, a peak-based centring algorithm is used instead.\footnote{By default, this uses the \texttt{photutils.centroids.centroid\_com} function, which determines the centroid as the intensity-weighted centre of mass.} In \texttt{EPSFBuilder}, this re-centring step is iterated over a small number of iterations. In practice, the definition of the ePSF centre is arbitrary provided it is applied consistently throughout the ePSF construction and fitting processes.

\paragraph*{Normalisation.} 
In \citet{Anderson_King_2000}, the ePSF is normalised such that a star with unit flux, centred exactly on the middle of a detector pixel, produces an image whose summed pixel values equal one. This definition ensures that the ePSF represents the fraction of a star's total light falling into each pixel as a function of its sub-pixel position. In contrast, there is no normalisation applied to the ePSF in \texttt{EPSFBuilder}.\footnote{Although a normalisation method exists within the underlying \texttt{\_LegacyEPSFModel} class used, it is never called in the \texttt{EPSFBuilder} class since the \texttt{\_LegacyEPSFModel} class is implemented using its default parameter \texttt{normalize=False}.} The normalisation of the ePSF is, in principle, arbitrary and only affects the overall flux scale. For applications using a single ePSF model, the lack of explicit normalisation simply results in a constant scaling of all measured fluxes, which has no effect on relative photometry or astrometry. However, when multiple ePSF models are used---for example, to capture the spatial variation in the ePSF across the image---the relative scaling of each ePSF model will differ. In such cases, normalisation is necessary to ensure consistent scaling of flux measurements between different ePSF models.

\paragraph*{End-of-iteration requirement.} 
In \citet{Anderson_King_2000}, the iterative process continues until the measured stellar fluxes and positions no longer exhibit significant pixel-phase–dependent errors. In contrast, in \texttt{EPSFBuilder} the iteration proceeds until either the user-specified maximum number of iterations is reached, or the measured source positions change by less than a defined tolerance between successive iterations, indicating that the ePSF model has converged. With this requirement, it is recommended to validate that the final ePSF removes pixel-phase biases in the final measured fluxes and positions.

\paragraph*{Degeneracy-breaking.} 
In \citet{Anderson_King_2000}, both the fluxes and positions of stars observed across multiple frames are averaged in order to break the degeneracy between the intrinsic stellar flux and the apparent flux variations caused by non-uniform intra-pixel sensitivity. In contrast, in \texttt{EPSFBuilder} only the stellar positions are averaged between iterations, while the fluxes are treated independently for each image. This difference in the degeneracy-breaking step reduces the accuracy of the resulting ePSF model, as the pixel-phase–dependent flux variations are not fully corrected. This is particularly significant for undersampled images which are more affected by intra-pixel sensitivity variations, as discussed further in Section~\ref{subsec:degeneracy_breaking}.

\vspace{0.5\baselineskip}
In this work, we make a number of modifications to the \texttt{photutils.psf} subpackage to bring it more closely into alignment with the ePSF modelling routine described by \citet{Anderson_King_2000}. We also implement several extensions that allow the ePSF modelling to be performed using different interpolation functions, gridpoint estimation methods, smoothing kernels, and ePSF fitting functions. These modifications provide additional flexibility to the ePSF modelling and fitting routines which allow for more accurate ePSFs to be generated, as discussed in Section~\ref{sec:ePSF_building}. The changes and additions are described in detail in Appendix~\ref{ap:code_changes}. The modified code is available from the authors upon request and will subsequently be submitted for integration into the \texttt{photutils} package.

\section{Simulating the ePSF}
\label{sec:simulations}

This work uses simulated ePSFs and images to investigate the accuracy of the ePSF modelling described in Section~\ref{subsec:epsf_modelling}, and of the resulting photometric and astrometric measurements. Synthetic ePSFs are derived by integrating instrumental PSFs over a model for the intra-pixel sensitivity profile. 

We model each detector pixel as a square collecting area with uniform sensitivity, separated by narrow gaps of zero response between adjacent pixels (a \textit{flat with gaps} intra-pixel sensitivity profile). This geometry can be represented by a two-dimensional top-hat function. We quantify the size of the gaps by the fill factor (FF), which is defined as the ratio of a pixel's light sensitive area to its total area. With this model we can recover a uniform intra-pixel sensitivity profile (\textit{flat and flush}) by using a full fill factor (FF $=1$). This work aims to to demonstrate the ability of the ePSF modelling procedure to fully account for a given non-uniform intra-pixel sensitivity profile in the ePSF model. Although simplified, the \texttt{flat with gaps} model is sufficient to demonstrate the principal effects of non-uniform intra-pixel sensitivity on the ePSF.

For simplicity, we adopted Gaussian forms for the instrumental PSF and explored three representative cases:
\begin{itemize}
    \item A simple circular Gaussian,
    \item A rotated elliptical Gaussian, and
    \item A skewed Gaussian.
\end{itemize}

The derivations of the synthetic ePSFs used to generate the simulated images are provided in Appendix~\ref{ap:analytic_epsfs}.

Figure~\ref{fig:input_epsfs} shows three synthetic ePSFs corresponding to the cases above, computed for pixels with a \textit{flat and flush} sensitivity profile. Later in this work, we use synthetic ePSFs computed for pixels with small gaps between them in order to investigate the effect of non-uniform intra-pixel sensitivity profiles.

\begin{figure}
    \centering
    \includegraphics[width=\columnwidth]{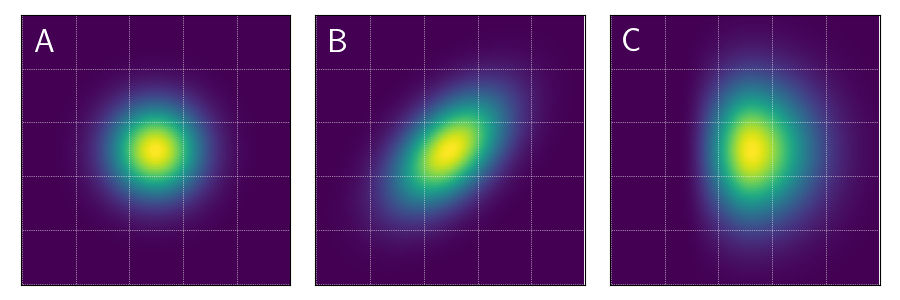}
    \caption{Synthetic ePSFs constructed by integrating a PSF model over `flush and flat' intra-pixel sensitivity profile. A: Circular Gaussian PSF with $\sigma = 0.5$ pixels. B: Rotated elliptical Gaussian PSF with $\sigma_x = 0.8$ pixels and $\sigma_y = 0.4$ pixels rotated by $45^\circ$. C: Skewed Gaussian with $\sigma_x = \sigma_y = 0.8$ and skew parameter $\eta = 5$. Dashed lines show pixel boundaries.}
    \label{fig:input_epsfs}
\end{figure}

\begin{figure}
    \centering
    \includegraphics[width=\linewidth]{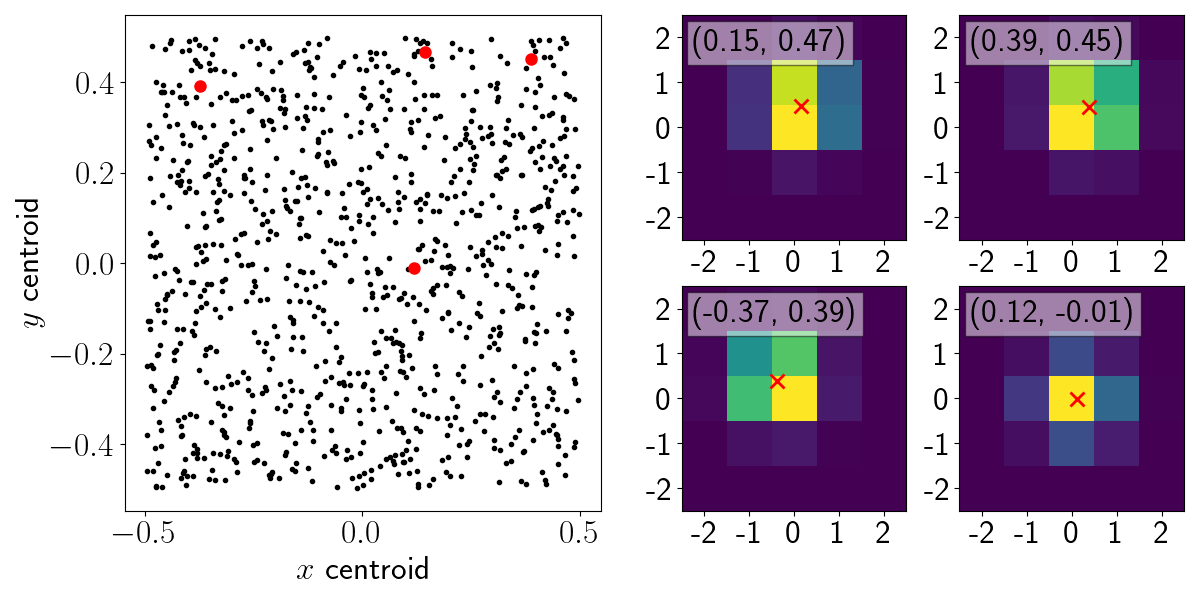}
    \caption{Left: Sub-pixel positions of 1000 random centroid coordinate samplings used to generate star images. Right: Examples of $5 \times 5$ pixel star images generated using synthetic ePSF A in Fig \ref{fig:input_epsfs}, with the numbers indicating their input sub-pixel centroid positions.}
    \label{fig:simulated_stars}
\end{figure}

The synthetic ePSF is evaluated over a grid of pixels with a given centroid position and flux value to generate a star image. This is repeated for a random sample of centroid positions and fluxes to produce a simulated dataset of star images as shown in Figure \ref{fig:simulated_stars}. We can use these simulated stars to generate an empirical ePSF model and compare against the known input ePSF. In addition, we can then perform ePSF photometry on these simulated stars using the generated empirical ePSF model and compare the results with their input positions and fluxes. This simulated framework forms the basis for our analysis in Section~\ref{sec:ePSF_building}.

\section{Accuracy of ePSF modelling}
\label{sec:ePSF_building}

For ePSF fitting photometry, the accuracy of the photometric measurements depends directly on how well the ePSF itself is modelled. Any imperfections in the model can introduce systematic pixel-phase errors in the measured fluxes and positions of stars. Constructing an accurate ePSF model, however, is a complex process influenced by multiple instrumental and (for ground-based telescopes) atmospheric factors, summarised below.

\begin{itemize}
    \item \textbf{PSF sampling.} When the image is undersampled, meaning the light from a star falls on only a small number of pixels, each star provides limited information about the PSF shape. This makes the ePSF harder to constrain and increases systematic errors.
    
    \item \textbf{Intra-pixel sensitivity.} Non-uniform pixel response causes the measured fluxes and positions of stars to depend on their sub-pixel location, introducing pixel-phase systematics in aperture photometry. These effects must be accounted for during ePSF modelling to prevent them from being imprinted on the ePSF model.
    
    \item \textbf{Geometric distortion.} This distortion alters the mapping between the true sky positions of stars and their measured detector coordinates. If uncorrected, it introduces systematic errors into the ePSF model. Accurate distortion correction is therefore essential for precise ePSF modelling \citep{Anderson_King_2003, Anderson_King_2004}.
    
    \item \textbf{Position dependence.} Optical effects can cause the PSF to vary across the detector. If these spatial variations are not included in the ePSF modelling, the resulting model will be inaccurate in different regions of the image. This can be mitigated by constructing a grid of ePSFs \citep{Anderson_King_2000}.
    
    \item \textbf{Brightness dependence.} In some detectors, particularly CCDs, charge from bright sources can repel subsequent charge, causing the ePSF to broaden for brighter stars \citep{Antilogus_2014}. This ``brighter–fatter'' effect means that a single ePSF model cannot accurately represent stars across all flux levels. This could be corrected by modelling the flux dependence of the ePSF or by generating separate ePSFs for different flux ranges.
    
    \item \textbf{Temporal variations.} The PSF can change over time due to telescope focus shifts, temperature fluctuations, mechanical flexure, or varying atmospheric conditions. If not tracked and modelled, these variations introduce inconsistencies into the ePSF, particularly in time-domain surveys or long observing campaigns. Temporal changes can be accounted for by applying small perturbations to the averaged ePSF model \citep{Anderson_King_2006} or by using a phylogram-based technique \citep{Anderson_Bedin_2017, Bellini_2018}.
\end{itemize}

In addition to instrumental and observational factors, the accuracy of the ePSF model depends on several user-defined parameters in the modelling routine that must be tuned to the specific implementation. The effect of each parameter on the accuracy of the ePSF model is examined in the following subsections.

In this work, we limit our analysis to the modelling of a \textit{stationary} ePSF --- one that does not vary with position, brightness, or time. While position-, flux-, and time-dependent variations are important to characterise for a complete description of an instrument’s behaviour, doing so first requires an accurate stationary ePSF. We therefore continue with this simplified assumption.

\subsection{Oversampling}
\label{subsec:oversampling}

The oversampling parameter defines the number of gridpoints per pixel in the ePSF model. For an oversampling $o$, there will be $o^2$ grid points per pixel. The optimal oversampling to use is dependent on the shape and size of the ePSF, and on the sample data available to construct the model. If the oversampling is too small, fine structure details are lost and interpolation errors increase; if it is too high, there may be insufficient sample sources to accurately constrain the ePSF model. 
An oversampling of four is commonly used in the literature, following its use for the WFPC2 camera of the HST in \cite{Anderson_King_2000}. This choice for the oversampling has been widely used across both space-based and ground-based instruments with diverse PSF characteristics \citep{Anderson_King_2004, Anderson_King_2006, Anderson_2006, Anderson_2015, Anderson_2016, Libralato_2014, Libralato_Feb2016, Libralato_Dec2016, Libralato_2023, Libralato_Mar2024, Libralato_Dec2024, Nardiello_2019, Nardiello_2022}. However, it is unlikely that a fixed oversampling of four is optimal for all instruments, as the best oversampling should depend on the true ePSF morphology. The repeated use of this default value may be due to the difficulty of knowing the optimal oversampling without already knowing the detailed ePSF structure.

To isolate the effect of oversampling on ePSF accuracy, we assume perfect knowledge of the ePSF values at the gridpoints. Rather than estimating them from simulated star samples, we evaluate the ePSF directly at the $(\Delta x, \Delta y)$ of the gridpoints from the synthetic input ePSFs described in Section~\ref{sec:simulations}. The resulting ePSF model, reconstructed from these tabulated values, can then be compared directly to the true synthetic ePSF to assess the interpolation error.
The ePSF model is then used to fit the fluxes and positions of 1000 simulated stars generated with the same synthetic input ePSF. Comparing the measured and true fluxes and positions allows us to quantify ePSF fitting accuracy and the scale of pixel-phase errors. We fix input fluxes to be unity, so the flux residual $f_\mathrm{fit} - f_\mathrm{input}$ is also the fractional error; position residuals are given in pixels as $x_\mathrm{fit} - x_\mathrm{input}$.

\begin{figure*}
	\includegraphics[width=\textwidth]{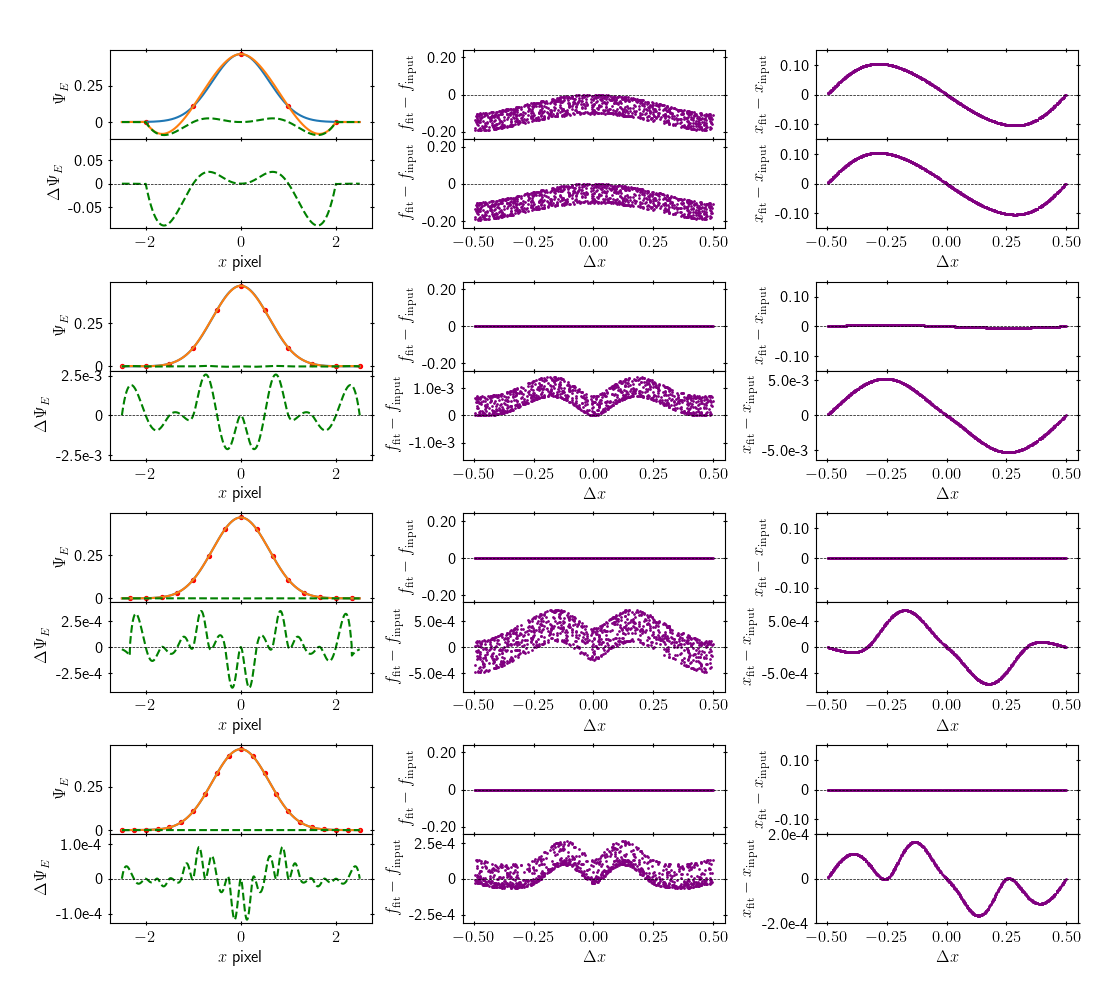}
    \caption{ePSF modelling and fitting accuracy of synthetic ePSF~A, comparing different oversampling parameters. Rows top to bottom are: oversampling $=1, 2, 3$, and 4). Left: True ePSF (blue) and interpolated ePSF model (orange) at $y=0$; residuals ($\Delta \Psi_E =$ interpolated - true ePSF) are shown as green dashed line with a magnified view in the lower panel. Red points mark the tabulated ePSF values used in the interpolation. Middle: Pixel-phase bias in fitted fluxes of 1000 simulated stars. $\Delta x$ is the $x$-displacement of the centroid of the star from the centre of a pixel. Since $f_\mathrm{input} = 1$, the flux residual $\Delta f = f_\mathrm{fit} - f_\mathrm{input}$ represents the fractional flux error. Right: Pixel-phase bias of the fitted $x$ centroid positions, with residuals $x_{\text{fit}} - x_{\text{input}}$ in pixel units.}
    \label{fig:oversampling_RectBivariateSpline}
\end{figure*}

\begin{figure}
    \centering
    \includegraphics[width=\linewidth]{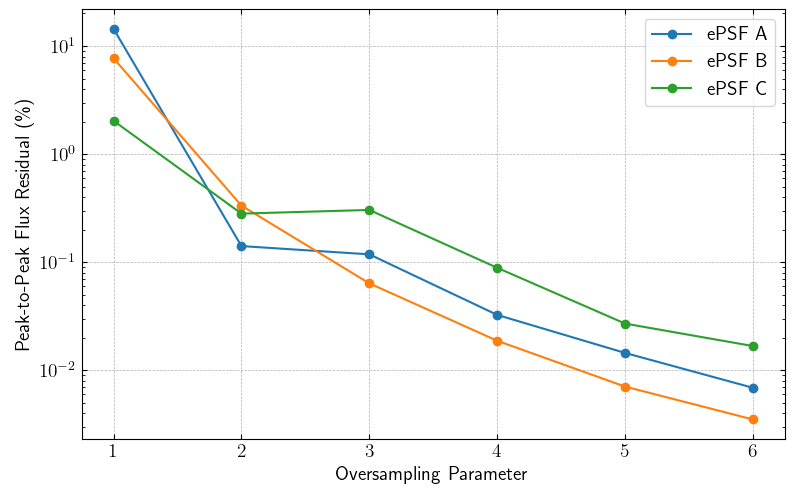}
    \caption{Pixel-phase error amplitude in flux measurements as a function of oversampling for the three synthetic ePSFs in Figure~\ref{fig:input_epsfs}. The amplitude is derived from the peak-to-peak variation in the median flux residuals across sub-pixel phase.}
    \label{fig:oversampling_summary}
\end{figure}

We first consider the simplest case--- a circular Gaussian PSF and a \textit{flat and flush} intra-pixel sensitivity profile (ePSF Model~A in Figure \ref{fig:input_epsfs}). Figure \ref{fig:oversampling_RectBivariateSpline}, compares oversampling parameters of $o=1$ -- 4. As anticipated, the interpolated model over-fits for low oversampling parameters. In the lowest sampling case of $o=1$, this leads to large pixel-phase errors, reaching amplitudes of up to $\sim$20 per cent in flux and $\sim$0.3 pixels in centroid position. Increasing the oversampling significantly reduces both interpolation artefacts and pixel-phase errors. 
We repeat these experiments for the other two synthetic ePSFs shown in Figure~\ref{fig:input_epsfs} and for a wider range of oversampling parameters. The scale of the pixel-phase errors in the flux measurements is quantified using a two-dimensional pixel-phase map of the median fractional flux residuals in $(\Delta x, \Delta y)$. Each cell of this pixel-phase map represents the median residual for stars whose position centroids fall within that sub-pixel region. The amplitude of the pixel-phase effect is then defined as the difference between the maximum and minimum cell medians, providing a measure of the peak-to-peak systematic variation in the measured fluxes. The resulting pixel-phase amplitudes for the different oversampling parameters are shown in Figure~\ref{fig:oversampling_summary}. Since we have assumed perfect ePSF values at the gridpoints, these measurements represent a lower limit to the pixel-phase errors expected in real data, where uncertainties in the gridpoint estimation introduce additional scatter and bias.

In the ePSF-building routine, the oversampling factor defines the number of gridpoints per detector pixel and is therefore independent of the intrinsic size of the ePSF. Consequently, for the same oversampling, a well-sampled ePSF that spans many pixels will be modelled more accurately than an undersampled one whose light is concentrated in fewer pixels. In practice, one should use the highest oversampling that the available sample of stars can reliably constrain (see Section~\ref{subsec:star_sample_number} for the dependence on sample size); when the oversampling is limited by the number of star samples, a minimum acceptable oversampling can be estimated from the requirement to keep pixel-phase errors negligible.
Our simulations show that for the approximately Gaussian synthetic ePSF (Model~A), an oversampling factor of three results in photometric and astrometric pixel-phase errors on the order of 0.1\%. We measure the full width at half maximum (FWHM) of this ePSF to be about 1.35~pixels.\footnote{The FWHM of the ePSF is broader than that of the instrumental PSF due to the integration over pixel areas.} This corresponds to roughly four gridpoints per FWHM. As a rule of thumb, we therefore recommend choosing an oversampling that provides at least four gridpoints per FWHM for near-Gaussian ePSFs. For non-circular ePSFs, the FWHM should be measured along the narrowest dimension, and for strongly asymmetric ePSFs, it should be estimated from the region with the steepest spatial gradient.

\subsection{Interpolation}
\label{subsec:interpolation}

To fit the empirical ePSF model to a star image, the ePSF must be evaluated at positions between the tabulated gridpoints. This is typically achieved using a third-degree bivariate spline interpolator to estimate the ePSF value at arbitrary $(\Delta x, \Delta y)$ coordinates. As shown in the previous section, interpolation errors can introduce significant pixel-phase biases in the measured fluxes and positions, particularly when the ePSF is sampled with only a few gridpoints per FWHM. The interpolation method must therefore be flexible enough to capture genuine variations in the ePSF between gridpoints, but not so flexible that it over-fits the tabulated values. We repeat the experiment from Figure~\ref{fig:oversampling_summary} for a range of alternative interpolation methods, summarised in Table~\ref{tbl:interpolators}. We again compare the amplitude of the pixel-phase errors on the flux measurements, estimated from the peak-to-peak flux residuals. 

The two best-performing interpolation methods across all three synthetic ePSF models (A--C) are the third-degree bivariate spline (RBS $k=3$) and the cubic radial basis function (RBF cubic). At low oversampling factors ($o=1$--3), the Gaussian RBF also performs well, though its accuracy plateaus and ultimately degrades relative to other methods at higher oversampling ($o\geq4$). For Models~A and~B, the RBS $k=3$ interpolator performs poorly in the most undersampled case ($o=1$), producing the largest pixel-phase errors of any method. In nearly all cases, the RBF cubic interpolator yields slightly smaller residuals than the RBS $k=3$ interpolator, maintaining superior accuracy across a range of ePSF shapes and sampling levels. We therefore adopt the RBF cubic interpolator for all subsequent analysis. This is implemented using the \texttt{RBFInterpolator} class from the SciPy library (\cite{SciPy_2020}).

\begin{table*}
    \centering
    \begin{tabular}{| m{4cm} | m{5cm} | m{3cm} |}
        \hline
        \textbf{Interpolator Label} & \textbf{Description} & \textbf{Kernel Function}\\ 
        \hline
        RBS $k=2$ & 2\textsuperscript{nd} degree bivariate spline approximation. & \\ 
        \hline
        RBS $k=3$ & 3\textsuperscript{rd} degree bivariate spline approximation. & \\ 
        \hline
        RBF thin\_plate\_spline & Radial basis function interpolation with a thin plate spline kernel. & $\phi(r) = r^2 \ln(r)$\\
        \hline
        RBF cubic & Radial basis function interpolation with a cubic kernel. & $\phi(r) = r^3$\\
        \hline
        RBF multiquadric $\epsilon=1.5$ & Radial basis function interpolation with a multi-quadric kernel with $\epsilon=1.5$. & $\phi(r) = \sqrt{(r/\epsilon)^2+1}$\\
        \hline
        RBF inverse\_multiquadric $\epsilon=1.0$ & Radial basis function interpolation with an inverse multi-quadric kernel with $\epsilon=1.0$. & $\phi(r) = \frac{1}{\sqrt{(r/\epsilon)^2+1}}$\\
        \hline
        RBF inverse\_quadratic $\epsilon=1.0$ & Radial basis function interpolation with an inverse quadratic kernel with $\epsilon=1.0$. & $\phi(r) = \frac{1}{(r/\epsilon)^2+1}$\\
        \hline
        RBF Gaussian $\epsilon=1.0$ & Radial basis function interpolation with a Gaussian kernel with $\epsilon=1.0$. & $\phi(r) = \exp(-(r/\epsilon)^2)$\\
        \hline
    \end{tabular}
    \caption{Labels and descriptions of interpolator methods tested for the interpolation of the ePSF model between gridpoint values. These are implemented in Python using the SciPy library \citep{SciPy_2020}.}
\end{table*}
\label{tbl:interpolators}

\begin{figure*}
    \centering
    \includegraphics[width=\textwidth]{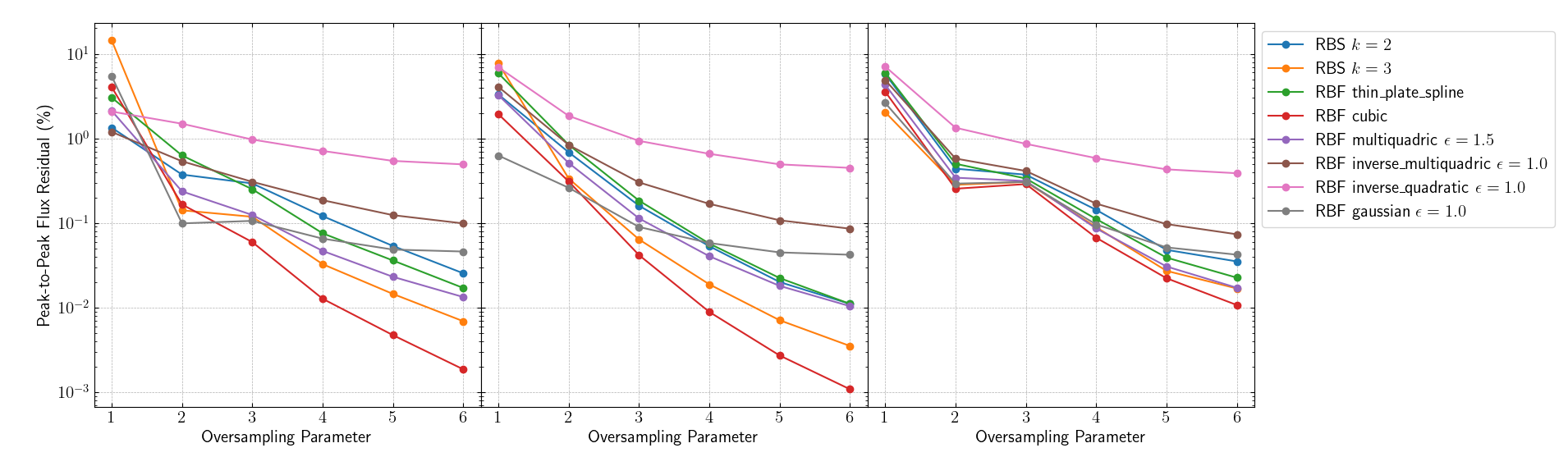}
    \caption{Comparison of different interpolation methods for ePSF fitting photometry with the synthetic ePSF models A (left), B (middle) and C (right) shown in Figure \ref{fig:input_epsfs}, across a range of oversampling values. Each circular point represents the pixel-phase error amplitude in the flux measurements of 1000 simulated sources, for the given oversampling, interpolator and input model.}
    \label{fig:interp_methods}
\end{figure*}

\subsection{Gridpoint Value Estimation}
\label{subsec:gridpoint_estimation}

\begin{figure}
    \centering
    \includegraphics[width=0.8\linewidth]{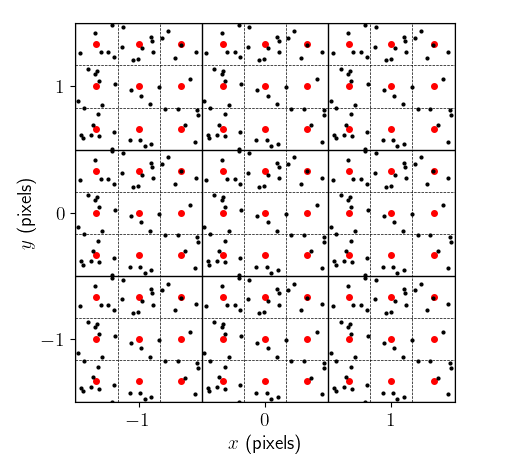}
    \caption{Demonstration of gridpoints and gridsections in the ePSF modelling for an oversampling factor of three. Solid black lines mark pixel boundaries, and dashed lines show the gridsection boundaries. Red points indicate the gridpoints, and black points show the locations of the ePSF samples from 50 simulated star images (each $3\times3$\,pixels) with randomly distributed sub-pixel centroids.}
    \label{fig:gridsection_gridpoint_demo}
\end{figure}

In the previous section, we assumed perfect knowledge of the ePSF values at each gridpoint. In practice, these values are estimated from the ePSF samples within small regions around each gridpoint, hereafter referred to as \textit{gridsections}. The standard approach is to take the sigma-clipped median of all ePSF samples that fall within each gridsection. The size of these regions is determined by the oversampling factor: for example, an oversampling of four divides each pixel into a $4 \times 4$ grid of sub-regions, with gridpoints located at their centres. Figure~\ref{fig:gridsection_gridpoint_demo} illustrates this geometry for a $3 \times 3$\,pixel ePSF model with an oversampling of three.

The accuracy of the sigma-clipped median for estimating ePSF gridpoint values depends strongly on the distribution of samples within each gridsection and on the shape of the ePSF. The sigma-clipped median will result in a bad estimate for the gridpoint values if the samples are clustered in a particular region of the gridsection. Furthermore, this method will always underestimate the gridpoint values for a gridsection which contains a local extrema. This is the case for the central gridpoint which is placed at the peak of the ePSF, where the sigma-clipped median estimation will always be smaller than the true value. 

\begin{figure*}
    \includegraphics[width=\textwidth]{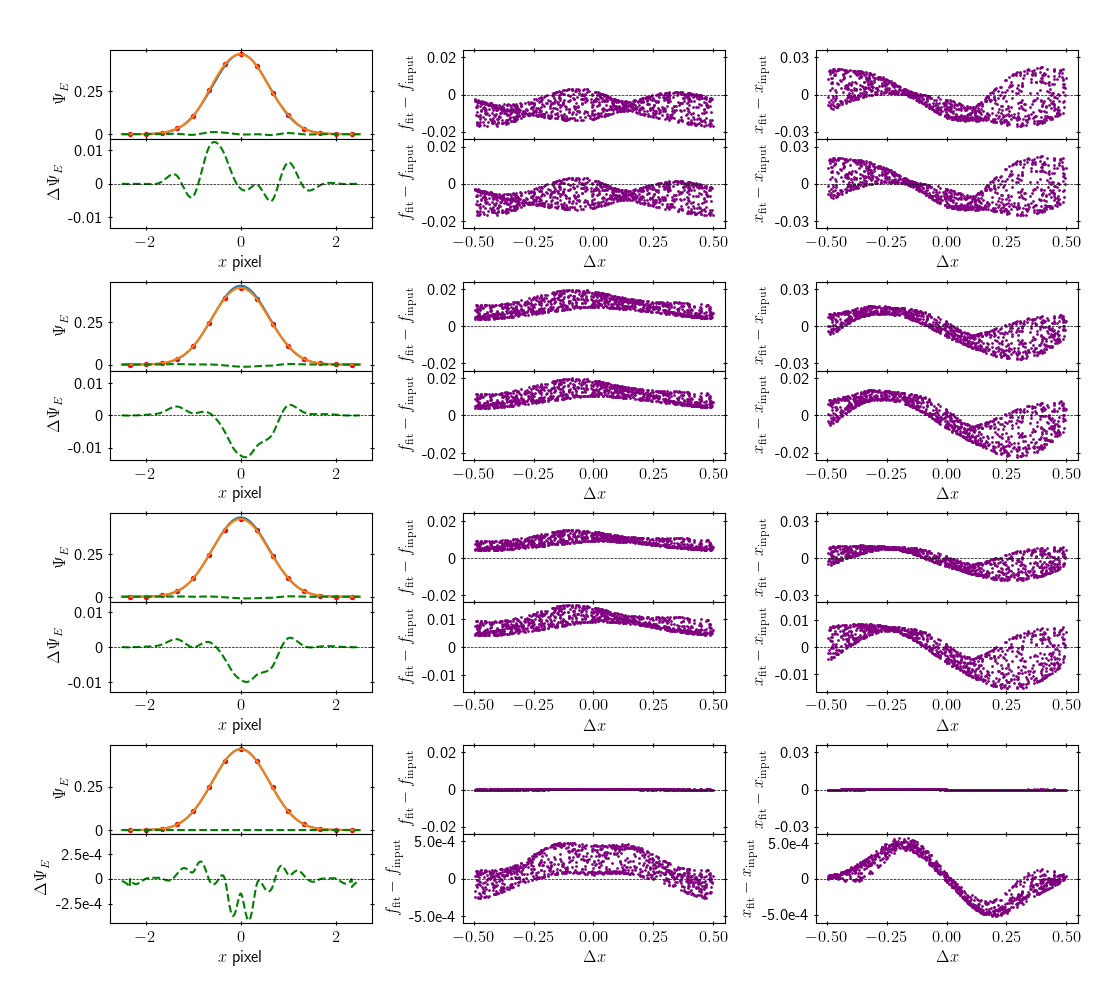}
    \caption{ePSF modelling and fitting accuracy of input ePSF Model A analogous to Figure \ref{fig:oversampling_RectBivariateSpline}. We model the ePSF with 1000 sample stars, use an oversampling parameter of 3, and use the RBF cubic interpolator method. Here we compare different methods for the gridpoint value estimation, where rows from top to bottom are: sigma-clipped median, sigma-clipped mean, sigma-clipped weighted mean, and 2D polynomial surface fitting.}
    \label{fig:gridpoint_estimation_o3_n1000}
\end{figure*}

\begin{figure*}
    \includegraphics[width=\textwidth]{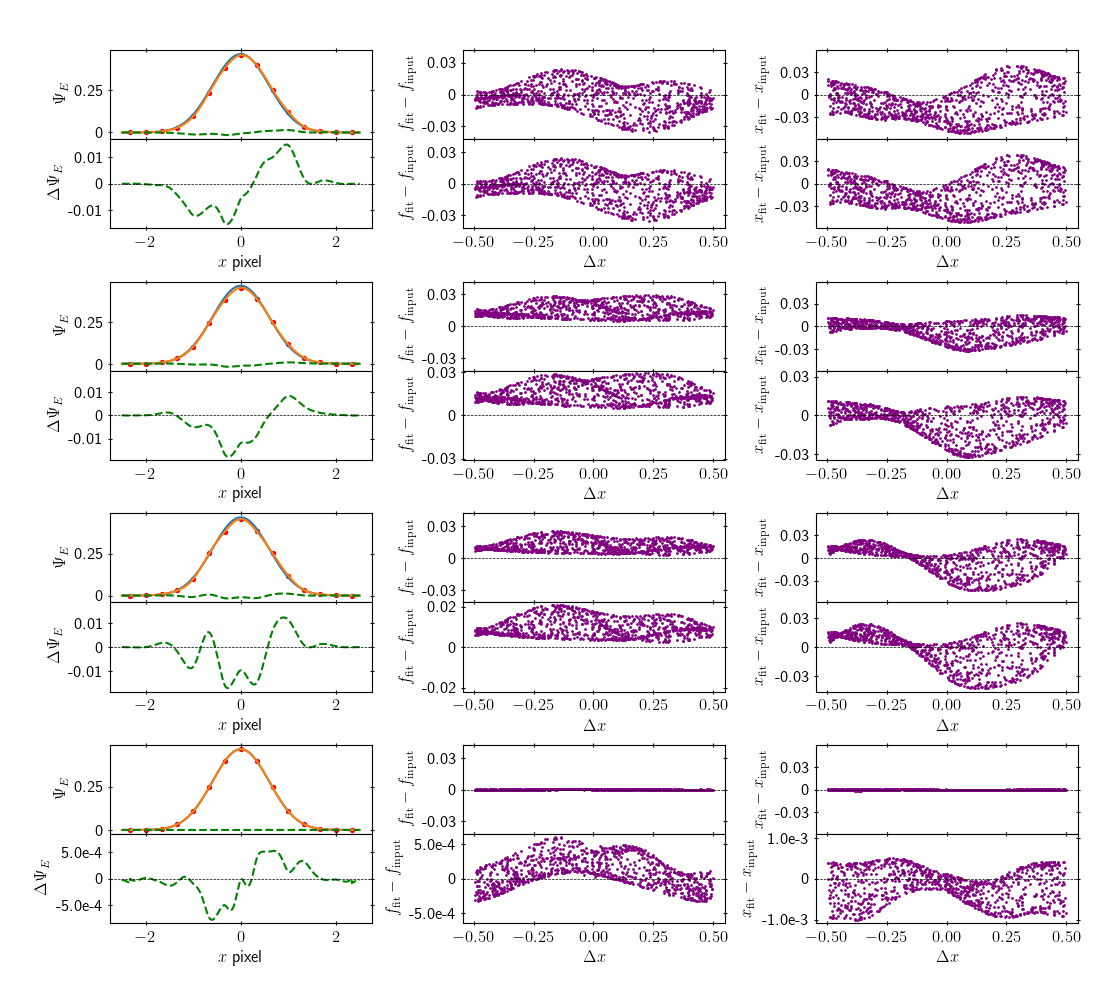}
    \caption{ePSF modelling and fitting accuracy of input ePSF Model A analogous to Figure \ref{fig:oversampling_RectBivariateSpline}. We model the ePSF with 100 sample stars, use an oversampling parameter of 3, and use the RBF cubic interpolator method. Here we compare different methods for the gridpoint value estimation, where rows from top to bottom are: sigma-clipped median, sigma-clipped mean, sigma-clipped weighted mean, and 2D polynomial surface fitting.}
    \label{fig:gridpoint_estimation_o3_n100}
\end{figure*}

Alternative methods are investigated for estimating the gridpoint values from the samples within the grisections: sigma-clipped mean, sigma-clipped weighted mean, and a 2D polynomial surface fitting. For the weighted mean, samples are weighted by their distance from the gridpoint position. For the polynomial method, we fit a two-dimensional polynomial surface over the samples within a gridsection. The value of the gridpoint is then estimated by evaluating the fitted surface at the gridpoint location. The polynomial used is either linear or quadratic depending on the number of sources within the gridsection. We generate 1000 simulated star images using the synthetic ePSF Model~A to construct ePSF models with each of the gridpoint estimation methods. The known fluxes and centroid positions of the simulated stars are used so that differences in the resulting models arise solely from the gridpoint estimation technique. A second, independent set of 1000 simulated star images is then used to test the photometric accuracy of each ePSF model (Figure~\ref{fig:gridpoint_estimation_o3_n1000}). We repeat the experiment using only 100 sample stars for the ePSF modelling to examine how inhomogeneous sample distributions affects each method (Figure~\ref{fig:gridpoint_estimation_o3_n100}). In both cases, the two-dimensional polynomial surface-fitting method produces residuals more than an order of magnitude smaller than the alternatives, yielding significantly more accurate photometry. This approach estimates the central gridpoint value more accurately and is less sensitive to uneven sample distributions within the gridsections.

\begin{figure*}
    \includegraphics[width=\textwidth]{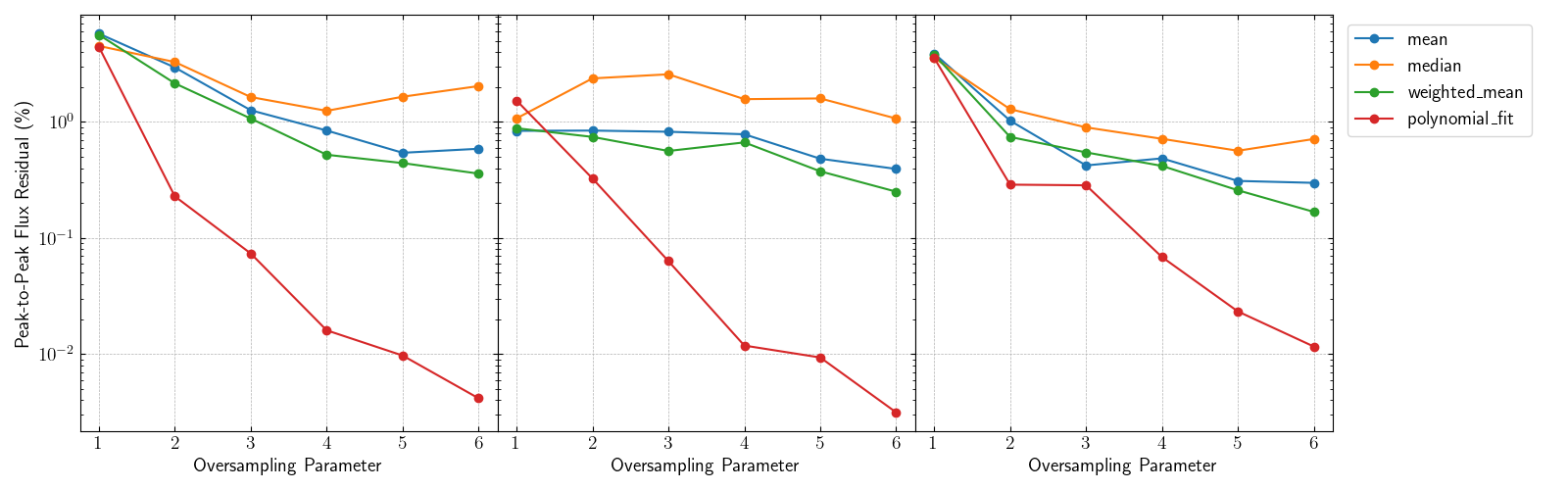}
    \caption{Comparison of different gridpoint estimation methods for ePSF fitting photometry for synthetic ePSF models A (left), B (middle), and C (right) as shown in Figure \ref{fig:input_epsfs}, across a range of oversampling values. Each circular point represents the pixel-phase error amplitude in the flux measurements of 1000 simulated sources, for the given oversampling, gridpoint estimation method, and input model.}
    \label{fig:gridpoint_estimation_summary}
\end{figure*}

We extend this analysis to a range of oversampling factors and to all three synthetic ePSFs shown in Figure~\ref{fig:input_epsfs}. Figure~\ref{fig:gridpoint_estimation_summary} compares the amplitude of the pixel-phase errors in the flux measurements of 1000 simulated stars for each input model, oversampling value, and gridpoint estimation method. Across all input ePSF models (A-C), the two-dimensional polynomial surface-fitting method consistently yields the smallest pixel-phase errors for nearly all oversampling factors. The amplitude of these errors decreases by several orders of magnitude at higher oversampling, largely due to the polynomial method’s ability to accurately estimate the ePSF at the central (peak) gridpoint, where other methods tend to underestimate the true value. At oversampling factors of $o\geq3$, the residuals from all estimation methods exceed those expected from interpolation errors alone (cf.\ Figure~\ref{fig:interp_methods}), indicating that gridpoint estimation uncertainty becomes the dominant source of error at these sampling levels. For the remainder of this work, the two-dimensional polynomial surface-fitting method is used to determine the gridpoint values in the ePSF modelling.

\subsection{Smoothing}
\label{subsec:smoothing}

As described in the previous section, the empirical ePSF model consists of estimated ePSF values at discrete gridpoints. In practice, these estimates are derived from noisy sample star images, introducing noise into the gridpoint values. To mitigate this, \citet{Anderson_King_2000} imposed a smoothness constraint on the ePSF model by convolving it with a smoothing kernel. In their implementation, with an oversampling factor of four, the kernel is a $5\times5$ least-squares quartic kernel with coefficients:

\begin{equation}
\begin{bmatrix}
  0.041632 & -0.080816 & -0.078368 & -0.081816 & 0.041632 \\
 -0.080816 & -0.019592 & 0.200816 & -0.019592 & -0.080816 \\
  0.078368 & 0.200816 & 0.441632 & 0.200816 & 0.078368 \\
 -0.080816 & -0.019592 & 0.200816 & -0.019592 & -0.080816 \\
  0.041632 & -0.080816 & -0.078368 & -0.081816 & 0.041632
\end{bmatrix}.
\end{equation}\label{eqn:standard_quartic_kenel}

The appropriate smoothing kernel depends on the oversampling factor (number of gridpoints per pixel), the intrinsic size and shape of the ePSF, and the level of noise in the gridpoint values. The kernel adopted by \citet{Anderson_King_2000} may not be optimal for all instruments or data qualities, so alternative kernels should be explored. In this section, we investigate how kernels of different sizes and polynomial degrees affect the accuracy of the ePSF modelling and the resulting ePSF-fitting photometry.  

We construct smoothing kernels by fitting a two-dimensional polynomial to an array of the desired kernel size, where the central element is set to one and all others are set to zero. To ensure a well-defined centre, the kernel dimensions are required to be odd. The resulting kernel is convolved with the empirical ePSF model, which is defined in oversampled coordinates. Consequently, the physical scale of the kernel relative to the detector pixels depends on the ePSF oversampling factor---for example, a $7\times7$ kernel applied to an ePSF with an oversampling of 7 corresponds to a kernel scale of one pixel, however the same kernel applied to an ePSF with an oversampling of three corresponds to a kernel scale of 2.33 pixels.  

We generate ePSF models for the synthetic ePSF Model~A from simulated star images following the procedure described in the previous section. To simulate realistic observational conditions, photon shot noise is added to the star images. For each pixel, the observed value is drawn from a Poisson distribution with a mean equal to the noiseless model intensity, representing the statistical uncertainty in photon counting. For bright sources, this is well approximated by adding Gaussian noise with a standard deviation equal to the square root of the pixel value. We set the noiseless intensity of each star to be $10^5$ counts. A constant background of 100 counts per pixel and Gaussian read noise with an RMS of 5 counts were also included. We use a smaller sample of 250 star images in this experiment to reproduce conditions where the noise on star images results in noisy gridpoint estimations. The resulting noisy star images are then used to construct the empirical ePSF models using the different smoothing kernels tested in this section. The independent set of 1000 simulated stars used to evaluate the ePSF-fitting accuracy is kept noise-free, ensuring that any differences in the recovered fluxes arise solely from imperfections in the ePSF model rather than noise in the fitting sample.

\begin{figure*}
    \includegraphics[width=\textwidth]{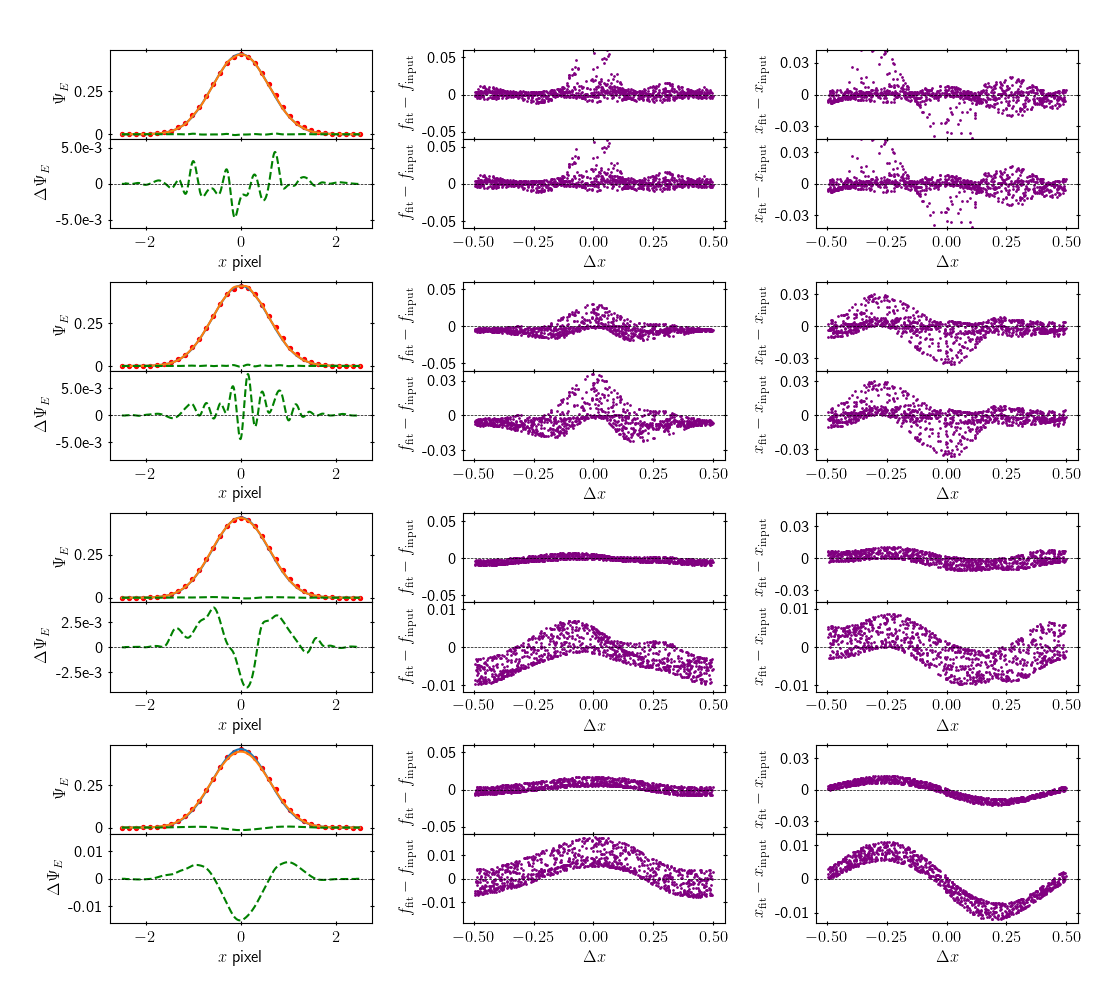}
    \caption{ePSF modelling and fitting accuracy of input ePSF Model A analogous to Figure \ref{fig:oversampling_RectBivariateSpline}. Here we fix the oversampling to 7 samples per pixel and compare the ePSF models with different smoothing kernels. Rows from top to bottom: no smoothing kernel, $5\times5$ quartic kernel, $9\times9$ quartic kernel, and $13\times13$ quartic kernel.}
    \label{fig:smoothing_kernel_size}
\end{figure*}

Figure~\ref{fig:smoothing_kernel_size} shows the effect of the smoothing kernel size on the ePSF modelling, compared with an un-smoothed model. Each kernel is constructed using a quartic polynomial. The $5\times5$ kernel produces residuals of similar amplitude to the un-smoothed case, indicating that a kernel of this size is insufficient to suppress the random noise in the gridpoints. Conversely excessively large kernels ($13\times13$) over-smooth the ePSF, erasing real small-scale structure in the model which results in large errors on the ePSF model and large pixel-phase errors.

\begin{figure*}
    \includegraphics[width=\textwidth]{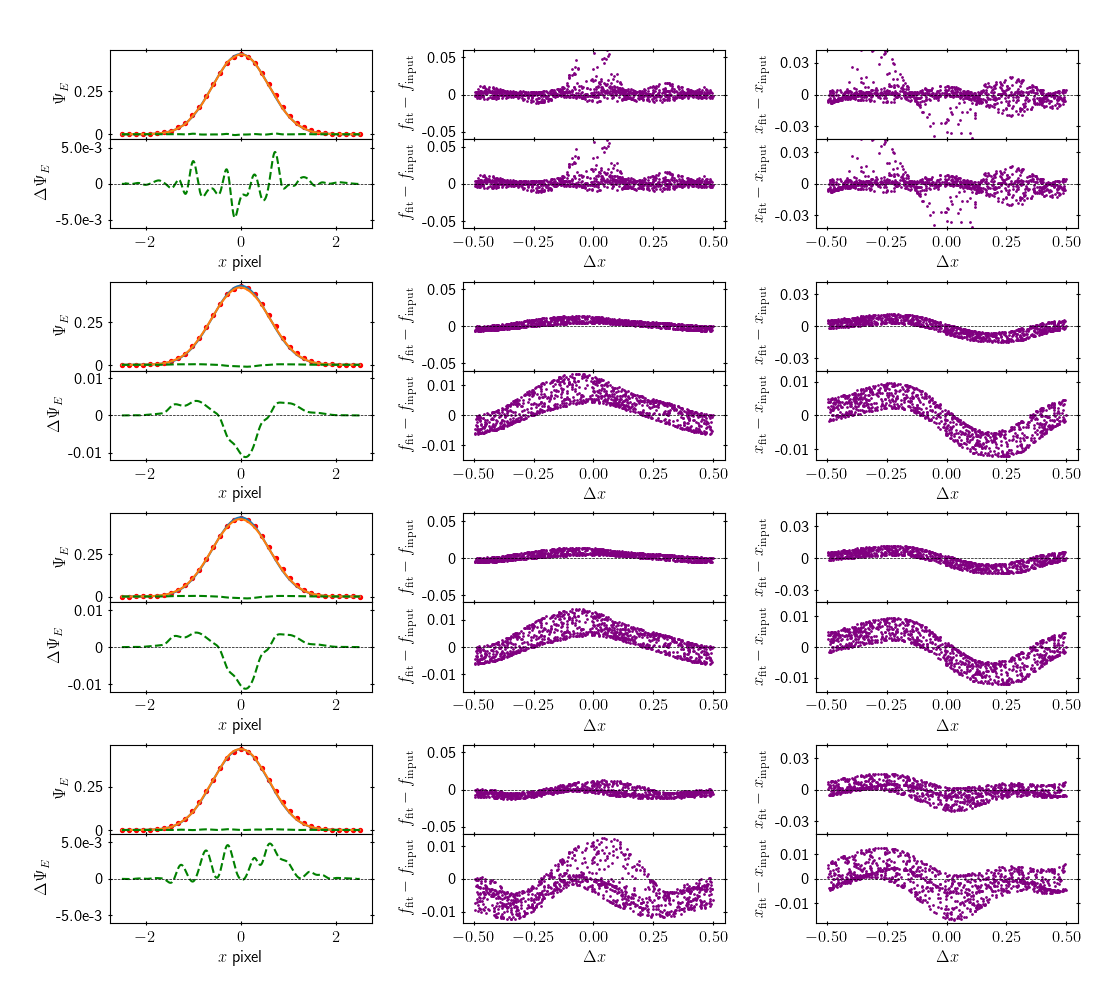}
    \caption{ePSF modelling and fitting accuracy of input ePSF Model A analogous to Figure \ref{fig:oversampling_RectBivariateSpline}.  Here we fix the oversampling to 7 and compare the ePSF models with different smoothing kernels. Rows from top to bottom: no smoothing kernel, $7\times7$ quadratic kernel, $7\times7$ cubic kernel, and $7\times7$ quartic kernel.}
    \label{fig:smoothing_kernel_function}
\end{figure*}

Next, we investigate the effect of the polynomial degree used to generate the smoothing kernel. Figure~\ref{fig:smoothing_kernel_function} shows results from a similar experiment to that in Figure~\ref{fig:smoothing_kernel_size}, but with the kernel size fixed at $7\times7$ (equivalent to $1\times1$\,pixels for $o=7$). We compare quadratic, cubic, and quartic polynomial kernels. In this case, the three kernels produce ePSF models with comparable residuals and pixel-phase errors. Both the quadratic and cubic models show evidence of oversmoothing, whereas the quartic kernel introduces oscillatory artifacts.

When the smoothing kernel is too large, it flattens the ePSF model and introduces significant pixel-phase errors in the photometry and astrometry. This effect is most pronounced for models with small oversampling factors, where the kernel spans a larger fraction of a pixel (e.g. a $7\times7$ kernel at $o=3$ corresponds to $2.3\times2.3$,pixels). 
These experiments demonstrate that smoothing must be chosen with care. The optimal kernel depends on the oversampling factor, and on the ePSF’s intrinsic size and structure. For a moderate oversampling value ($\geq 3$), a range of smoothing kernel sizes should be tested in order to find the minimum smoothing kernel which removes the smoothing noise but does not cause over-smoothing problems. In the case of undersampled ePSF with smaller oversampling values, one should avoid smoothing unless there are large amounts of noise in the ePSF gridpoints.

\subsection{Number of Star Samples}
\label{subsec:star_sample_number}

In practice, one of the main limitations in ePSF modelling is the number of stellar samples available to constrain the model. The required sample size depends on how finely we want to super-sample the ePSF (the oversampling factor), since a higher oversampling---and therefore a larger number of gridpoints---requires more samples. For the WFPC2 ePSF, \citet{Anderson_King_2000} used 45,000 star samples (3000 per image across 15 images). To account for spatial variation across the field, they constructed nine independent ePSF models at evenly spaced detector positions, each based on roughly 5000 stars. These 5000 stars were plenty sufficient to estimate the ePSF at the 16 gridpoints per pixel for an oversampling of four. In contrast, ground-based applications of the ePSF routine use far fewer stars---typically 50--150 for the same oversampling factor \citep{Anderson_2006}. There is currently no clear guideline for how many star samples are required to achieve a given level of ePSF accuracy for a specific oversampling value.

It is clear that a sufficient number and distribution of stars is required to ensure that each gridsection contains enough ePSF samples to estimate the gridpoint values using the methods described in Section~\ref{subsec:gridpoint_estimation}. One way to achieve a well-distributed sampling is through a precisely dithered set of exposures, in which the sub-pixel centroids of stars are systematically shifted to cover all gridsections. Dithered images are also essential for mitigating the effects of intra-pixel sensitivity variations on the ePSF model, as discussed further in Section~\ref{subsec:dither_pattern}. However, in many practical cases such a dataset may not be available, and we must instead rely on the random distribution of stars to provide an adequate set of ePSF samples.

We perform a series of Monte Carlo simulations to determine how many randomly distributed star samples are required so that the probability of having at least $M$ ePSF samples per gridsection exceeds 70, 80, 90, or 95~per~cent for a given oversampling factor. The results for various $M$ and oversampling values are summarised in Table~\ref{tbl:n_sample_probs}. For the commonly used oversampling of four, we find that about 240 star samples provide a 95~per~cent probability of having at least six ePSF samples per gridsection—the minimum needed to fit a two-dimensional quadratic surface in the gridpoint estimation. Conversely, using fewer than about 60 stars reduces the probability of having even a single ePSF sample per gridsection below 70~per~cent. When no ePSF samples fall within a gridsection, the gridpoint value is estimated from surrounding points using either two-dimensional cubic spline or nearest-neighbour interpolation. This will add an additional interpolation error to the ePSF modelling which should be taken into account when using the model for photometric measurements. 

\begin{table}
    \centering
    \begin{tabular}{|c|c|c|c|c|c|c|}
        \hline
        Oversampling & Probability & \( M = 1 \) & \( M = 3 \) & \( M = 6 \) & \( M = 10 \)\\
        \hline
        \multirow{4}{*}{2} 
        & 70\% & 8 & 21 & 36 & 56\\
        & 80\% & 10 & 22 & 39 & 59\\
        & 90\% & 12 & 25 & 42 & 63\\
        & 95\% & 15 & 29 & 48 & 68\\
        \hline
        \multirow{4}{*}{3} 
        & 70\% & 28 & 56 & 97 & 142\\
        & 80\% & 32 & 63 & 102 & 150\\
        & 90\% & 39 & 72 & 112 & 164\\
        & 95\% & 44 & 78 & 121 & 172\\
        \hline
        \multirow{4}{*}{4} 
        & 70\% & 60 & 116 & 187 & 274\\
        & 80\% & 68 & 125 & 199 & 288\\
        & 90\% & 77 & 139 & 219 & 311\\
        & 95\% & 88 & 156 & 238 & 327\\
        \hline
        \multirow{4}{*}{5} 
        & 70\% & 105 & 196 & 311 & 454\\
        & 80\% & 119 & 210 & 328 & 472\\
        & 90\% & 136 & 237 & 358 & 501\\
        & 95\% & 163 & 265 & 383 & 539\\
        \hline
    \end{tabular}
    \caption{Minimum number of star samples, $N$, required to achieve the given probability of having $M$ samples per gridsection. Number of gridsections is determined by the oversampling as $o^2$.}
    \label{tbl:n_sample_probs}
\end{table}

\begin{figure*}
    \includegraphics[width=\textwidth]{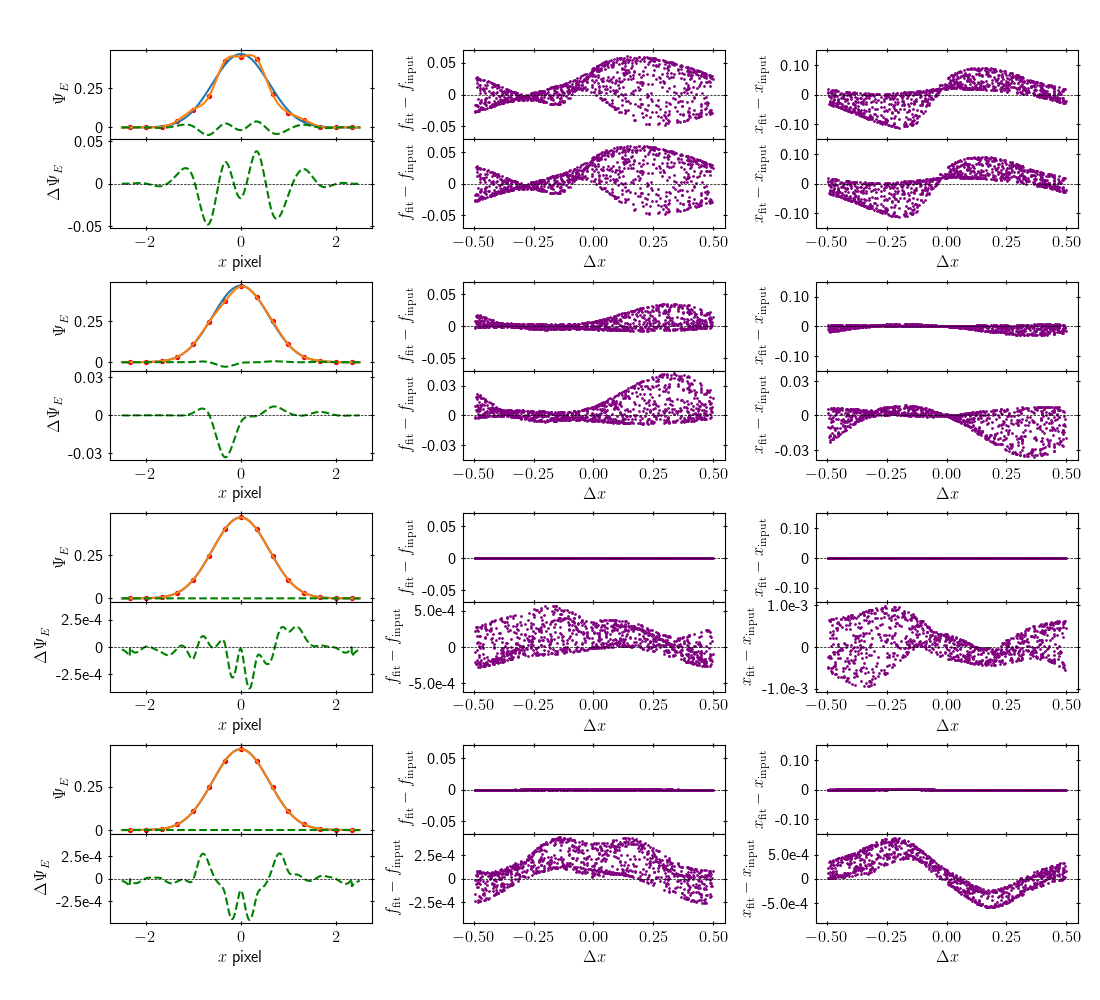}
    \caption{ePSF modelling and fitting accuracy of input ePSF Model A analogous to Figure \ref{fig:oversampling_RectBivariateSpline}. We fix the oversampling parameter to $o=3$ and test the effect of the number of star samples. Rows top to bottom are: 10, 50, 100, and 500 samples.}
    \label{fig:sample_number_o3}
\end{figure*}

\begin{figure*}
    \includegraphics[width=\textwidth]{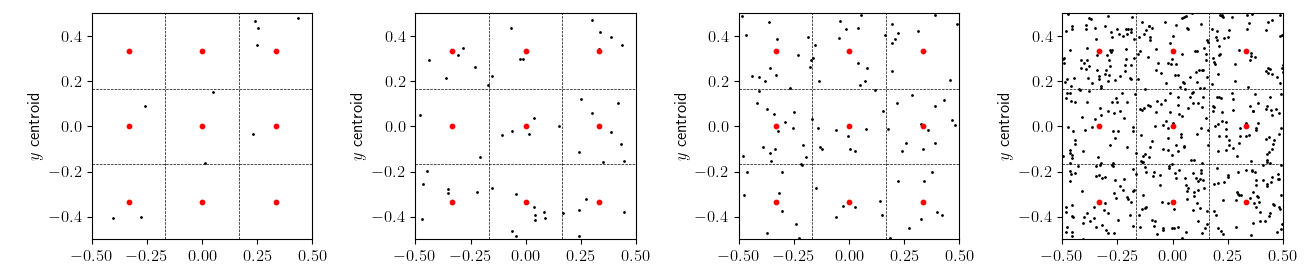}
    \caption{Distribution of sub-pixel centroid positions for different numbers of randomly distributed stars. Black points show the sub-pixel centroid positions. From left to right: 10, 50, 100, and 500 stars. Red points show the positions of gridpoints and dashed lines show the gridsection boundaries for an oversampling parameter of $o=3$.}
    \label{fig:sample_number_dist}
\end{figure*}

We test the effect of the star sample number, $N$, on the accuracy of the ePSF model generated following the experiments in Section~\ref{subsec:gridpoint_estimation}. 
Figure \ref{fig:sample_number_o3} shows the results for the synthetic ePSF Model A, using an oversampling of three, with randomly distributed star samples of 10, 50, 100, and 500. The corresponding sub-pixel distributions of centroid positions of the stars are shown in Figure~\ref{fig:sample_number_dist}. 
The distribution of the 10 samples is very inhomogeneous across the sub-pixel space, meaning some gridpoints are evaluated by just one sample, or must be interpolated from neighbouring points. If a gridpoint is evaluated by just one sample, the single sample could be from anywhere within the $1/3 \times 1/3$ pixel gridsection. the estimated ePSF values often deviate substantially from the true values, producing large residuals and significant pixel-phase errors in both flux and position. 
Increasing the sample to 50 stars reduces these errors but still leaves some under-sampled regions, as evident in Figure~\ref{fig:sample_number_dist} (e.g. the middle-left gridsection contains only two samples).
For $N=100$ and $N=500$, the sample distribution becomes more uniform, resulting in much more accurate ePSF reconstructions. The corresponding photometric and astrometric pixel-phase errors are several orders of magnitude smaller than in the cases with fewer star sampling, confirming that an adequate and homogeneous sample is essential for reliable ePSF modelling.

\begin{figure}
    \includegraphics[width=\linewidth]{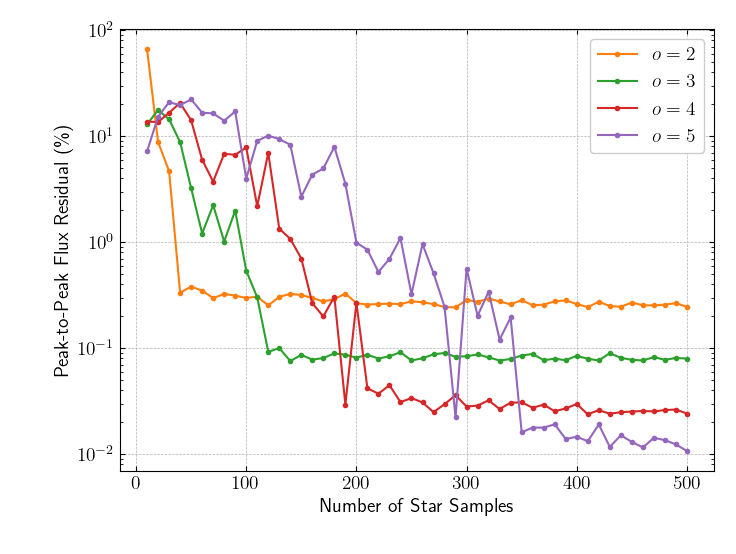}
    \caption{Comparison of the effect of the number of sample stars used in the ePSF modelling on the ePSF fitting photometry results for synthetic ePSF Model A shown in Figure \ref{fig:input_epsfs}. Each circular point represents the pixel-phaser error amplitude in the flux measurements of 1000 simulated stars for the given oversampling and number of sample stars.}
    \label{fig:sample_number_summary}
\end{figure}

We perform an experiment to estimate the minimum number of star samples required to accurately model the ePSF for a given oversampling factor. For the synthetic ePSF model A in Figure~\ref{fig:input_epsfs}, we generate ePSF models from simulated star samples while varying the number of stars between 10 and 550, repeating this for oversampling values of $o=2$--5. As in previous sections, we calculate the amplitude of the pixel-phase errors in the flux measurements from ePSF-fitting photometry of 1000 simulated stars, as shown in Figure~\ref{fig:sample_number_summary}.
For each oversampling factor, the pixel-phase error amplitude converges to a minimum when the number of stars approximately matches that predicted for a 95~per~cent probability of having at least $M=6$ samples per gridsection (Table~\ref{tbl:n_sample_probs}). For larger oversampling factors than those tested here, a similar Monte Carlo procedure can be used to estimate the number of stars needed for accurate ePSF modelling.
If precisely dithered images are available, this will reduce the total number of stars required to generate an accurate ePSF model.

\subsection{Degeneracy Breaking}
\label{subsec:degeneracy_breaking}

So far, we have not been considering the effect of intra-pixel sensitivity variations on the ePSF since we have been using synthetic ePSF models with a \textit{flat and flush} intra-pixel sensitivity profile. As shown previous sections, pixel-phase bias errors arise from having errors in the ePSF model, regardless of whether intra-pixel sensitivity variations are present or not. However, if intra-pixel sensitivity variations present, we will always end up with an incorrect estimate of the ePSF if we do not properly consider the degeneracy between the flux of stars and their sub-pixel centroid position. 

\begin{figure}
    \includegraphics[width=\linewidth]{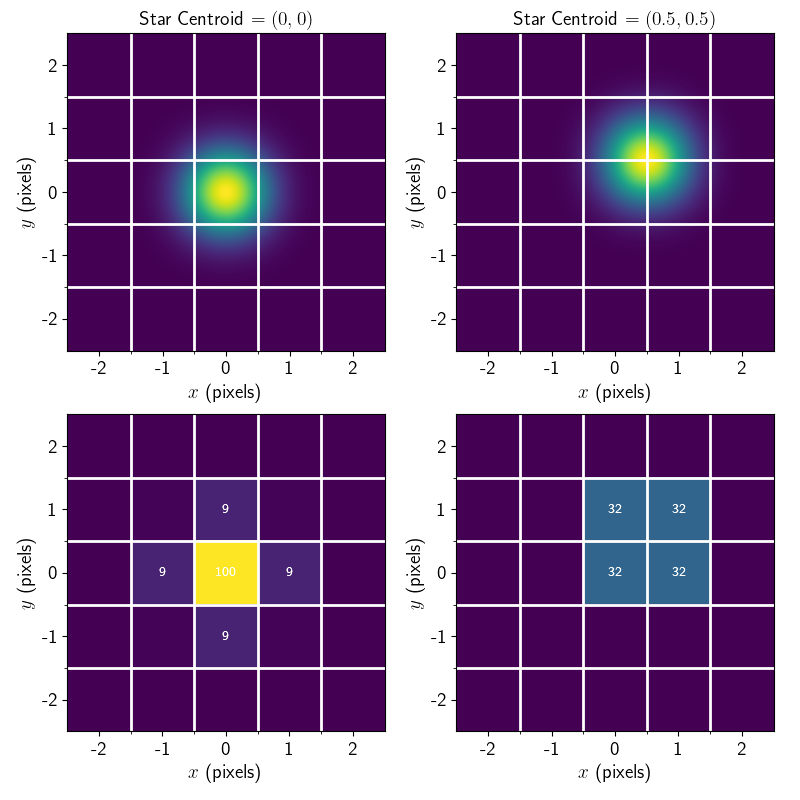}
    \caption{Top: The ePSF input model for a circular Gaussian PSF with $\sigma = 0.4$ and a \textit{flat with gaps} intra-pixel sensitivity profile (FF $= 0.64$). The model is evaluated for a star centred on a pixel (left) and for one centred at a pixel corner (right), both with the same input stellar flux. Bottom: Simulated stellar images corresponding to the two cases above. Solid lines indicate pixel boundaries, and numbers within the pixels show the simulated counts.}
    \label{fig:intra-pix_variability_demo}
\end{figure}

The response of a detector pixel to a photon depends on the exact location within the pixel where photon falls, for both CCD \citep{Lauer_1999} and CMOS \citep{Mahato_2018} sensors. This means the measured pixel values in a stellar image are dependent on both the true flux of the star and the positions of the pixels relative to the centroid of the star. To demonstrate this, Figure \ref{fig:intra-pix_variability_demo} shows simulated star images with identical input fluxes but different sub-pixel centroid positions. The input synthetic ePSF model used here has a circular Gaussian PSF ($\sigma=0.4$ pixels) and a \textit{flat with gaps} intra-pixel sensitivity profile (FF $=0.64$) which produces a non-uniform sensitivity that varies with sub-pixel position. The image generated for a star centred at the middle of a pixel has a higher aperture sum flux than the image of the star centred at the corner of a pixel, despite both images being generated with identical input fluxes. This difference in observed flux is entirely due to the non-uniform intra-pixel sensitivity. 

\begin{figure*}
    \includegraphics[width=\textwidth]{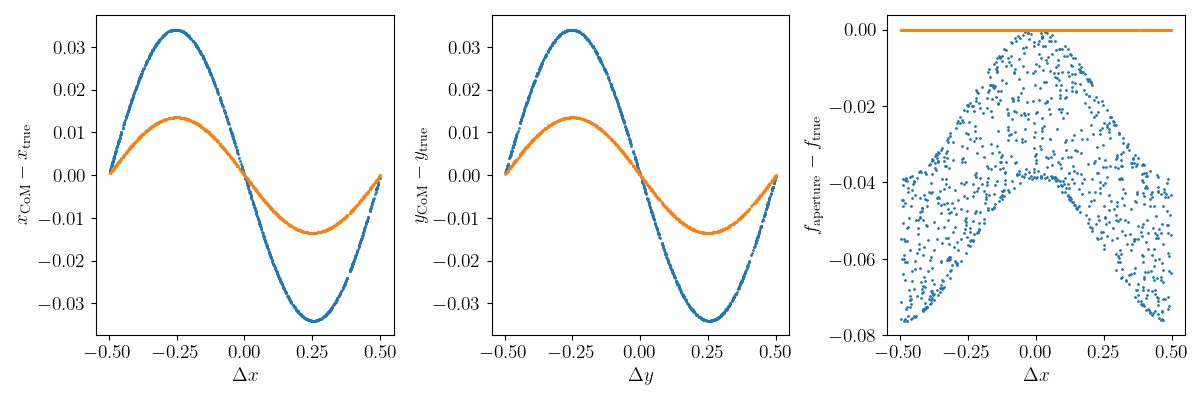}
    \caption{Centre-of-mass centroiding and aperture photometry measurements of 1000 simulated stellar images generated with an ePSF model for a circular Gaussian PSF ($\sigma = 0.4$) and two intra-pixel sensitivity profiles: \textit{flat-with-gaps} (FF $= 0.64$, blue) and \textit{flat-and-flush} (FF $= 1$, orange). Left: Pixel-phase errors in the $x$-centroid versus $x$ sub-pixel position. Middle: Pixel-phase errors in the $y$-centroid versus $y$ sub-pixel position. Right: Pixel-phase errors in flux versus $x$ sub-pixel position.}
    \label{fig:intra-pixel_PPE_demo}
\end{figure*}

If the stellar position and flux are measured directly from the pixel values (e.g. with centre-of-mass centroiding and aperture photometry), the results will depend on the star’s sub-pixel position. To demonstrate this, we generate 1000 star simulated images with the same input flux but at random sub-pixel positions for ePSFs with both \textit{flat with gaps} (FF $=0.64$) and \textit{flat and flush} (FF $=1$) intra-pixel sensitivity profiles. The centroid and flux of the star images are then measured with centre-of-mass centroiding and aperture photometry respectively. Figure \ref{fig:intra-pixel_PPE_demo} shows the resulting pixel-phase bias in the centroid and flux measurements of the simulated star images. A clear systematic trend is seen for the \textit{flat with gaps} model in both flux and centroid position. The \textit{flat and flush} model shows no flux bias, but does show a trend in the centroid position measurements due to inaccuracies in the centre-of-mass method for undersampled images, however this is smaller than for the \textit{flat with gaps} model.

Intra-pixel sensitivity variations affect our ability to construct the ePSF using the routine presented in \citet{Anderson_King_2000} (as described in Section \ref{subsec:epsf_modelling}). The first step in this process is to estimate the fluxes and positions of stars, for this we use a centre-of-mass centroid and aperture photometry. The initial flux measurement ($f^1_{\text{obs}}$) for a star with a true centroid position of $(x_*, y_*)$ is given by:

\begin{equation}
    f^1_{\text{obs}} = f_* \times \sum_{i,j} \Psi^{T}_{E}(i-x_*,j-y_*)
    \label{eqn:f1obs}
\end{equation}

\noindent where $f_*$ is the true flux of the star, $\Psi^T_E$ is the true ePSF of the instrument, and $i, j$ are integer values denoting which pixels are included in the aperture sum. The centre-of-mass centroid position, $(x^1_{\text{obs}}, y^1_{\text{obs}})$, is given by:

\begin{equation}
\begin{aligned}
    x^1_{\text{obs}} &= \frac{\sum_{i,j} i \times \Psi^T_E (i-x_*, j-y_*)}{\sum_{i,j} \Psi^T_E (i-x_*, j-y_*)}, \\
    y^1_{\text{obs}} &= \frac{\sum_{i,j} j \times \Psi^T_E (i-x_*, j-y_*)}{\sum_{i,j} \Psi^T_E (i-x_*, j-y_*)}.
\end{aligned}
\end{equation}

The next step is to use the flux and centroid estimates of stars and the individual pixel values, $P$, within the star images to construct a first model of the ePSF ($\Psi^1_E$) using the following relation:

\begin{equation}
    \Psi^1_E(i - x^1_{\text{obs}}, j - y^1_{\text{obs}}) = \frac{P(i - x^1_{\text{obs}}, j - y^1_{\text{obs}})}{f^1_{\text{obs}}} .
\end{equation}

The individual pixel values are determined by the true properties of the star and true ePSF, so we can re-write the above as:

\begin{equation}
    \Psi^1_E(i - x^1_{\text{obs}}, j - y^1_{\text{obs}}) = \frac{f_*\Psi^T_E(i - x_*, j - y_*)}{f^1_{\text{obs}}} .
\end{equation}

This ePSF model contains errors relating to both the errors on the centroid and flux estimates. Using Equation \ref{eqn:f1obs} for $f^1_{\text{obs}}$ and simplifying the notations for $(i-x_*, j-y_*)$ and $(i-x^1_{\text{obs}}, j-y^1_{\text{obs}})$ as $(\Delta x_*, \Delta y_*)$ and $(\Delta x^1_{\text{obs}}, \Delta y^1_{\text{obs}})$ respectively, we can write down the ratio between the model and true ePSF as:

\begin{equation}
    \frac{\Psi^1_E(\Delta x^1_{\text{obs}}, \Delta y^1_{\text{obs}})}{\Psi^T_E(\Delta x^1_{\text{obs}}, \Delta y^1_{\text{obs}})} = \frac{\Psi^T_E(\Delta x_*, \Delta y_*)}{\Psi^T_E(\Delta x^1_{\text{obs}}, \Delta y^1_{\text{obs}})\sum_{i,j} \Psi^{T}_{E}(\Delta x_*,\Delta y_*)}
    \label{eqn:epsf1_ratio} .
\end{equation}

The next step is to use this preliminary ePSF estimate to re-measure the stellar fluxes and positions through an iterative least-squares fitting procedure. Because this process is iterative, it is difficult to express the updated measurements of flux and position explicitly in terms of the true stellar values and the ePSF. To illustrate the effect of intra-pixel sensitivity variations on the ePSF modelling routine more clearly, we therefore simplify the problem by assuming that the true stellar positions are known, allowing Equation~\ref{eqn:epsf1_ratio} to be written as:

\begin{equation}\label{eqn:epsf_ratio_simplified}
    \frac{\Psi^1_E(\Delta x, \Delta y)}{\Psi^T_E(\Delta x, \Delta y)} = \frac{1}{\sum_{i,j} \Psi^{T}_{E}(\Delta x,\Delta y)},
\end{equation}

\noindent where we now have $\Delta x = \Delta x_* = \Delta x^1_{\text{obs}}$ and $\Delta y = \Delta y_* = \Delta y^1_{\text{obs}}$. 

This simplification removes the need for a least-squares fitting procedure to measure the stellar flux with the ePSF. Instead, the flux is simply the scaling factor between the pixel values of the stellar image and the corresponding ePSF values at the same pixel displacements. The measured flux for a star at position $(x_*, y_*)$ can therefore be written as:

\begin{equation}
    f^2_{\text{obs}} = \frac{P(\Delta x, \Delta y)}{\Psi^1_E(\Delta x, \Delta y)} = f_* \Psi^T_E(\Delta x, \Delta y) \times\frac{\sum_{i,j} \Psi^{T}_{E}(\Delta x,\Delta y)}{\Psi^T_E(\Delta x, \Delta y)} = f^1_{\text{obs}}
    \label{eqn:flux_no_dgcy_brk},
\end{equation}

\noindent where $(\Delta x, \Delta y)$ can be any $(i-x_*, j-y_*)$ for any $(i,j)$ pixel position within the star's image. We find that the recovered fluxes are consistent with those measured by aperture photometry. Any pixel-phase errors present in the initial flux measurements are propagated into the ePSF-derived fluxes through the corresponding errors introduced into the ePSF model. If we had not assumed the stellar positions were known in Equation~\ref{eqn:epsf_ratio_simplified}, additional pixel-phase dependent errors would arise through inaccuracies in the position measurements, further amplifying the pixel-phase biases in both the ePSF-derived fluxes and centroids. 

As described in \cite{Anderson_King_2000}, the issue with the ePSF modelling is due to the degeneracy between the flux of a star and its sub-pixel centroid position. They demonstrate that this degeneracy can be broken using dithered observations of the same stars. Instead of using the measured fluxes and positions of each individual star image to measure the ePSF, the fluxes and positions of each star are fixed to be their average across all dithered images---this is the key degeneracy breaking step highlighted in \cite{Anderson_King_2000}.






Continuing under the assumption the true positions of stars are known, the ePSF model constructed using stars with this fixed average flux, $\overline{f^2_{\text{obs}}}$, is given by:

\begin{equation}
    \Psi^2_E(\Delta x, \Delta y) = \frac{f_* \Psi^T_E (\Delta x, \Delta y)}{\overline{f^2_{\text{obs}}}} .
    \label{eqn:epsf_avg_flux}
\end{equation}


If an infinite number of dithered frames were available such that the stellar centroids uniformly sampled the sub-pixel space in both $x$ and $y$, the pixel-phase biases in the individual flux measurements would cancel out when averaged. In this ideal case, $\overline{f^2_{\text{obs}}}$ would equal the true stellar flux, $f_*$, the measured ePSF would be identical to the true ePSF, and subsequent flux measurements using this model would be free of pixel-phase errors.  

In practice, however, the number of dithered frames is finite, so we cannot always assume that $\overline{f^2_{\text{obs}}} = f_*$. The value of $\overline{f^2_{\text{obs}}}$ depends on both the dither pattern and the positions of the stars within the dithered images, approaching $f_*$ only for large numbers of uniformly distributed dithers. The effect of the dither pattern on breaking the flux--position degeneracy and producing accurate ePSF models is discussed further in Section~\ref{subsec:dither_pattern}.

Equation~\ref{eqn:flux_no_dgcy_brk} highlights the need to constrain both the position and the flux of stars in the dithered images to be their average values. We arrived at Equation~\ref{eqn:flux_no_dgcy_brk} under the assumption that we already knew the exact positions of stars, which has the same effect as if we had constrained the positions of the stars to their average value. However, using constrained star positions but not fluxes will still generate pixel-phase errors in the resulting flux measurements. 

\begin{figure*}
    \includegraphics[width=\textwidth]{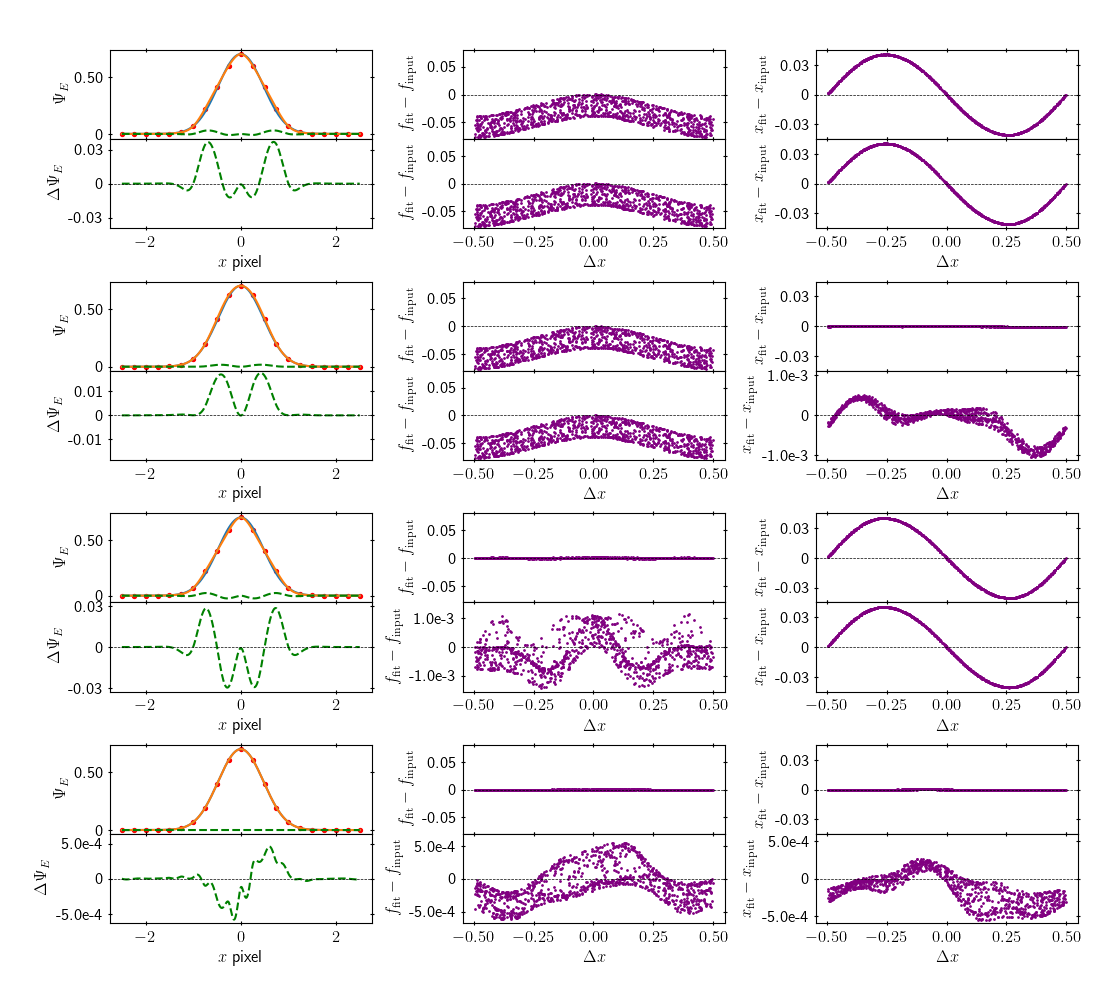}
    \caption{ePSF modelling and fitting accuracy using dithered images of input ePSF model for a circular Gaussian PSF with $\sigma=0.4$ and a \textit{flat with gaps} intra-pixel sensitivity profile (FF $=0.64$). Columns are analogous to Figure \ref{fig:oversampling_RectBivariateSpline}. The dither pattern used is a uniform dithering at $1/4$ pixel spacings. Rows from top to bottom: Neither flux nor positions constrained, only position constrained, only flux constrained, both position and flux constrained.}
    \label{fig:degeneracy_breaking}
\end{figure*}

We demonstrate this with our simulation framework in Figure~\ref{fig:degeneracy_breaking}. We evaluate the accuracy of the ePSF model and the resulting photometric and astrometric measurements on 1000 simulated star images under four different scenarios: neither flux nor position of stars are constrained, only position of stars are constrained, only flux of stars are constrained, both position and flux of stars are constrained. 
As expected, when no constraints are applied, strong pixel-phase errors appear in both the fitted fluxes and positions.
Constraining only the positions eliminates most of the bias on the positions but leaves flux errors unchanged, while constraining only the fluxes has the opposite effect.
When both quantities are constrained, pixel-phase errors in both flux and position are greatly reduced.

Users of the \texttt{photutils.psf} subpackage should note that by default, the \texttt{LinkedEPSFStar} class constrains only the positions of stars when linking dithered observations of the same source \citep{Bradley_2024}. As detailed in Appendix~\ref{ap:code_changes}, we modify the implemenation of the \texttt{photutils.psf} subpackage to constrain both fluxes and positions in the ePSF modelling. 

\subsection{Dither Pattern}
\label{subsec:dither_pattern}

As discussed in Section~\ref{subsec:degeneracy_breaking}, dithered observations are essential for breaking the degeneracy between stellar flux and sub-pixel position introduced by intra-pixel sensitivity variations. However, the effectiveness of this approach depends not only on the number of dithered frames but also on how well their sub-pixel positions sample the detector’s pixel phase space. In this section, we investigate how different dither patterns influence the accuracy of the ePSF modelling and the resulting photometric and astrometric measurements.

\begin{figure*}
    \includegraphics[width=\textwidth]{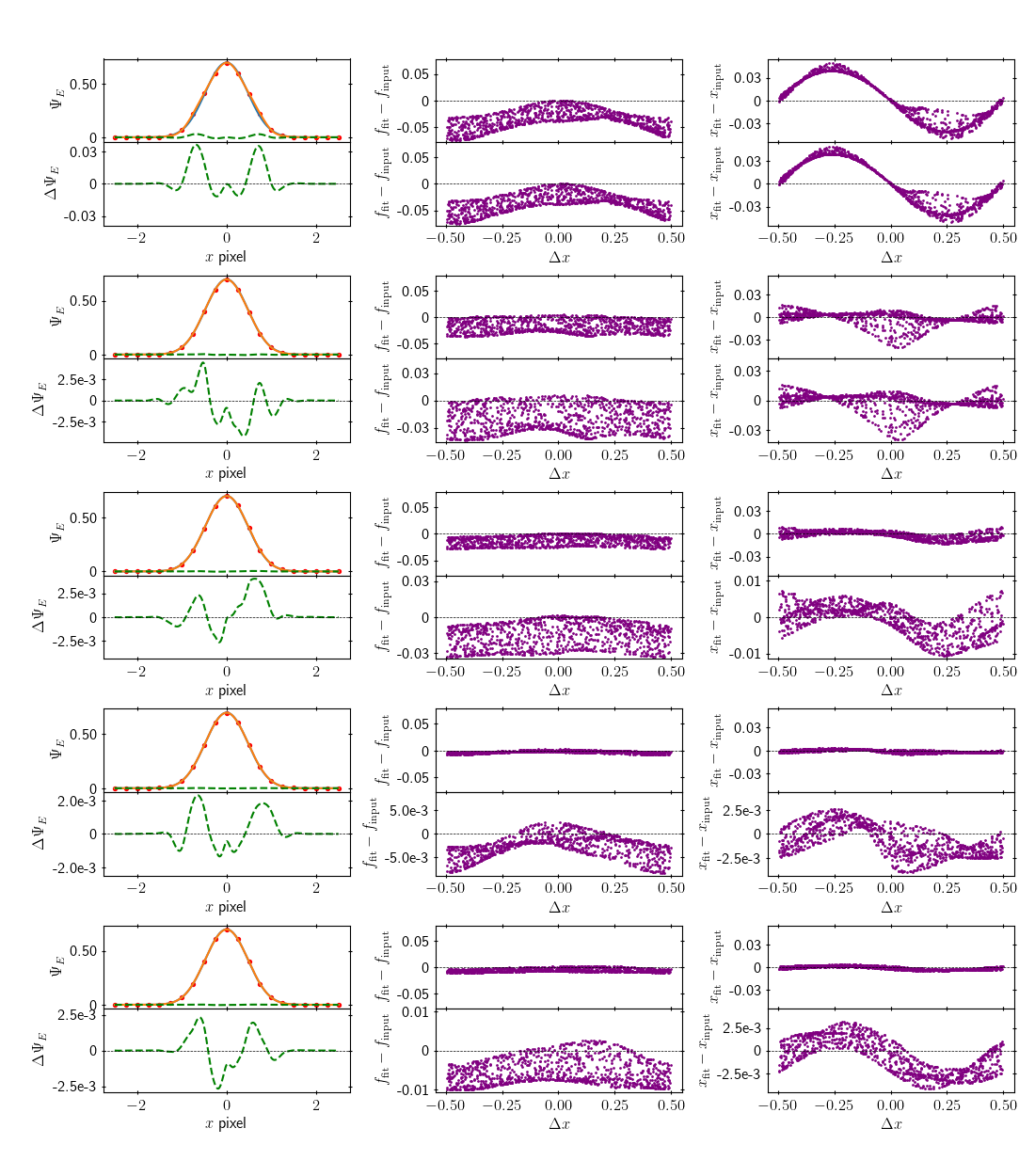}
    \caption{ePSF modelling and fitting accuracy of input ePSF model for a circular Gaussian PSF with $\sigma=0.4$ and a \textit{flat with gaps} intra-pixel sensitivity profile (FF $=0.64$). Columns are analogous to Figure \ref{fig:oversampling_RectBivariateSpline}. We investigate the effect of different dither patterns on the ePSF modelling. Rows from top to bottom: No dithering, 2 random ditherings, 4 random ditherings, 8 random ditherings, and 16 random ditherings.}
    \label{fig:intra-pix_dithers_random}
\end{figure*}

\begin{figure*}
    \includegraphics[width=\textwidth]{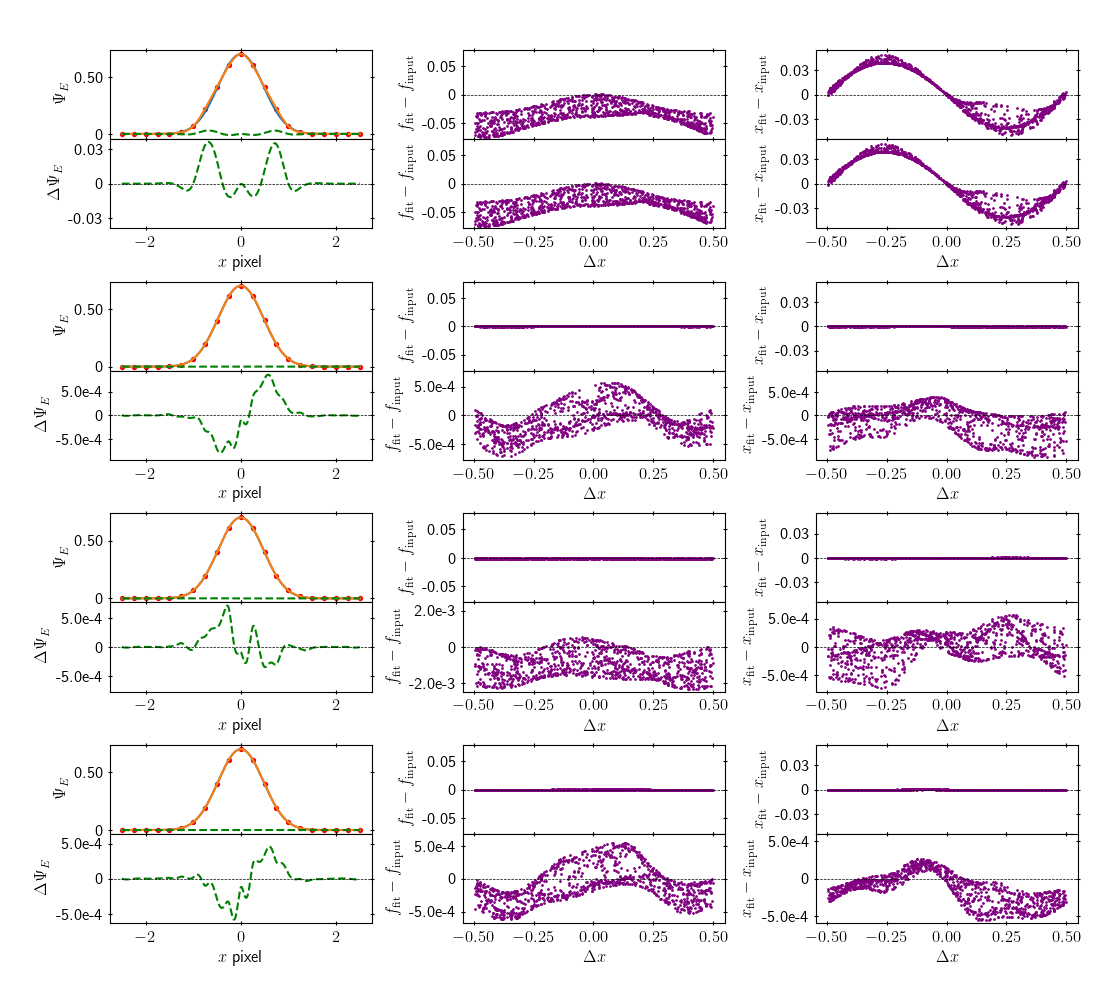}
    \caption{ePSF modelling and fitting accuracy of input ePSF model for a circular Gaussian PSF with $\sigma=0.4$ and a \textit{flat with gaps} intra-pixel sensitivity profile (FF $=0.64$). Columns are analogous to Figure \ref{fig:oversampling_RectBivariateSpline}. We investigate the effect of different dither patterns on the ePSF modelling. Rows from top to bottom: No dithering, uniform dithering at $1/2$ pixel spacings, uniform dithering at $1/3$ pixel spacings, and uniform dithering at $1/4$ pixel spacings.}
    \label{fig:intra-pix_dithers_spirals}
\end{figure*}

In Figures \ref{fig:intra-pix_dithers_random} and \ref{fig:intra-pix_dithers_spirals} we test the effect of different dither patters on the ePSF modelling for the same input ePSF model used in Figures \ref{fig:intra-pix_variability_demo} and \ref{fig:intra-pixel_PPE_demo}. For each dither pattern tested, we begin by generating 100 simulated star images with random centroid positions. We then make several copies of each star image where the centroids are shifted by small amounts corresponding to each of the dither positions for the given dither pattern. These star images are then used to construct the ePSF model with an oversampling of 4. In Figure \ref{fig:intra-pix_dithers_random}, we test random dither patterns corresponding to 2, 4, 8, and 16 randomly dithered images. In Figure \ref{fig:intra-pix_dithers_spirals}, we construct dither patterns which uniformly sample the sub-pixel space as $2 \times 2$, $3 \times 3$, and $4 \times 4$ grid of dither positions. In both cases, we compare the accuracy of the ePSF modelling and the resulting flux and position measurements to that for the case of no dithering. 

It is clear that a uniform dither pattern is preferable to random dithering. This is expected given the way that the ePSF is modelled as a set of values at specific gridpoints as discussed in Section \ref{subsec:gridpoint_estimation}. For example, say we have a dither pattern which matches the oversampling grid of the ePSF model. Each gridpoint is evaluated using the star ePSF samples which fall within the corresponding sub-pixel grid section. For such a dither pattern, each individual star will have its centre shifted in each image so that it falls within each gridsection in the central pixel. Therefore, each gridpoint in the ePSF model will be evaluated using samples from all of the stars, minimising the pixel-phase dependence on the ePSF model and resulting flux and position measurements of stars. Even with a dither pattern which doesn't fully match the oversampling grid, we still achieve much better results with the uniform dither patterns than random dither patterns with many more dithered frames. Therefore, when either planning dithered measurements or selecting dithered sample images to use for ePSF modelling, the focus should be on obtaining ditherings which uniformly sample the sub-pixel space rather than the number of dithered images obtained.

\section{Application to the Global Jet Watch}
\label{sec:GJW}

We now consider the application of the ePSF modelling routine described above to real astronomical images from the Global Jet Watch (GJW) commensal photometric cameras. The GJW is a network of observatories spread across the globe in longitude to enable round-the-clock spectroscopy and photometry. The GJW observatories were constructed with the goal of performing time-domain astronomy, whereby we investigate the varying nature of objects on timescales of hours and days. Each observatory is equipped with a 0.5m main telescope which collects light for a dual-arm fibre-fed spectrograph.
In addition, each observatory has three commensal photometry cameras which are attached to the frame of the main telescope, making use of the pointing, tracking and guiding of the main telescope. The three cameras allow us to observe in three different photometric filters: H-$\alpha$, H-$\beta$, and a continuum filter at 736nm. There have been two generations of these commensal cameras. The first generation took wide field images of $5.4 \times 4.3$ degrees on CCD detectors of size $2749 \times 2199$ pixels, while the second generation take images of $4.5 \times 4.5$ degrees on CMOS detectors of size $3008 \times 3008$ pixels. 
Since they are attached to the main telescope, the fields observed by the commensal cameras are determined by the particular objects being observed with the GJW spectrographs. The main scientific goal of these commensal cameras is to identify new transient objects, for instance new microquasar candidates, which will be followed-up with the GJW spectrographs. Full descriptions of the observatories and their instruments will be given in Blundell et al., in prep., and Lee et al., in prep.

The GJW commensal cameras have been collecting images since 2017, and our dataset currently consists of over 300,000 images. These images are processed by a bespoke photometric pipeline (full description of this pipeline will be given in Godden et~al., in prep). This pipeline generates lightcurves for each of the unique objects detected in the images. The current database consists of over 500,000 unique objects with each lightcurve potentially having tens of thousands of data points. A major challenge in the construction of the lightcurves is accurately measuring the stellar fluxes and positions. The GJW images are very undersampled, with the FWHM of our images being under two pixels, leading to significant pixel-phase systematics in lightcurves generated using aperture photometry. 

In this section, we investigate the use of the ePSF modelling and fitting routine described in this work to assess whether it is possible to reduce these pixel-phase systematics in the GJW lightcurves. We use the knowledge from previous sections to determine how to best tune the ePSF modelling routine to the GJW images, implementing this with our modified version of the \texttt{photutils.psf} subpackage. We compare our results with those from using the current release of the \texttt{photutils.psf} subpackage, and with the results from aperture photometry. 
As discussed in Section~\ref{sec:ePSF_building}, ePSF modelling and fitting with real data present additional challenges, including spatial variation of the ePSF across the detector, geometric distortion, the brighter–fatter effect, and temporal variations in atmospheric seeing. A full description of the ePSF modelling for the GJW cameras, accounting for these effects, will be presented in the second paper in this series, Godden et~al., in prep.
Here, we restrict our analysis to a small central region of the images, using stars within a narrow flux range and selecting frames with consistent atmospheric seeing, effectively treating the ePSF as spatially and temporally invariant. As a proof of concept, we present ePSF modelling results for the H$\alpha$-filter commensal camera (first generation CCD camera) at the South African GJW observatory (GJW-SA); the full set of ePSF models will be presented in Godden et~al., in prep.

\subsection{Sample Data}

To model the ePSF for the GJW-SA H-$\alpha$ camera, we use a set of images of the field around the classical nova V6595 Sgr. This nova was observed between the nights of the 17th July 2020 and the 20th July 2020 by the GJW spectroscopic system for an analysis on the precessing jets of classical novae \citep{McLoughlin_2021}. During the observation campaign of V6595 Sgr, the commensal photometric cameras obtained images of the field around V6595 Sgr with 100 second exposure times. After removing unsuitable images (e.g. with large cloud cover), there are a total of 237 images within this dataset to be used for ePSF modelling. The raw GJW-SA images were bias-, dark-, and flat-field corrected before being used for ePSF modelling. Details of the image calibration and background subtraction procedures are provided in Appendix~\ref{ap:gjw_preprocessing}.

To construct a stationary ePSF for the GJW-SA H-$\alpha$ data, we restrict the analysis to regions and stars where the PSF shape is approximately constant. The ePSF modelling is limited to the central $916\times732$~pixel region of the image, where the PSF is least distorted and geometric distortion effects are minimal. We further restrict the sample to stars with fluxes between $10^4$ and $10^5$~counts to mitigate brightness-dependent shape variations (the brighter--fatter effect). To minimise the impact of variable atmospheric seeing, only images with similar measured atmospheric seeing values are used, which limits our sample to 31 images. 

\begin{figure*}
    \centering
    \includegraphics[width=0.85\textwidth]{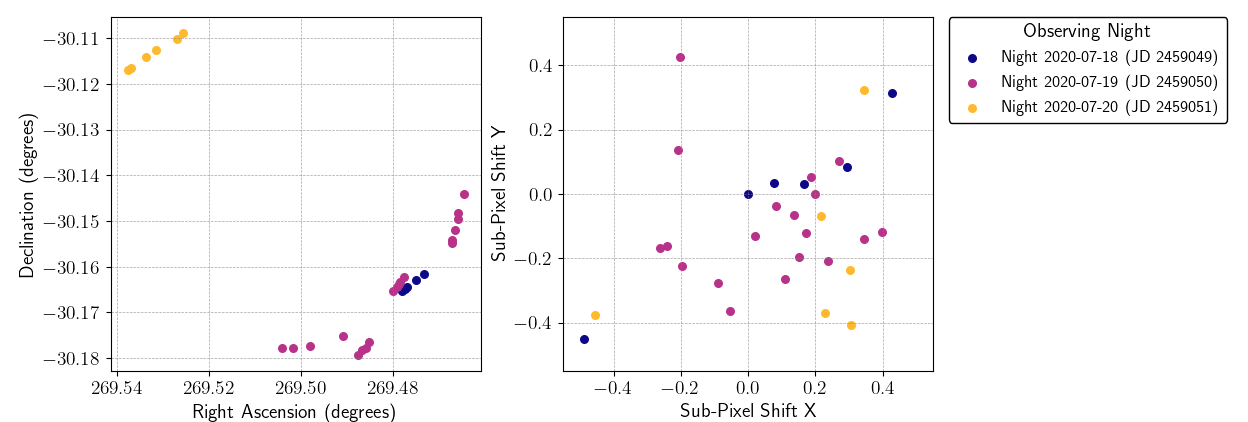}
    \caption{Dithering of sample images used in ePSF construction. Left: Variation in the pointing right ascension and declination of the images. Right: Sub-pixel displacements of the images with respect to the first image in the sample.}
    \label{fig:pointing_subpixel}
\end{figure*}

Figure~\ref{fig:pointing_subpixel} shows the dither pattern of this subset of the images. We show both the pointing right ascension and declination of the images (determined by the main telescope), and the sub-pixel displacements of the images. The sub-pixel displacements show how the sub-pixel centroids of each star in the images are varied compared to their position in the initial image of the sample. As discussed in Section~\ref{subsec:dither_pattern}, the optimum dither pattern would form a uniform grid of sub-pixel displacements. Although our sample does not exhibit a uniform pattern, the sub-pixel displacements are fairly well sampled in both the $x$ and $y$ directions. Given a sufficiently large set of sample of randomly distributed stars in each frame, this will be sufficient for ePSF modelling.

From these 31 images, we detect a total of 2,767 star samples (123 unique stars) within the specified position and flux ranges. We then remove star samples unsuitable for ePSF modelling, including: blended sources or stars with close neighbours that prevent clean cutouts; stars with noisy backgrounds and low signal-to-noise ratios; sources detected in only a few images; and stars showing flux variations inconsistent with pixel-phase effects, suggesting intrinsic variability. After applying these cuts, 2,515 star samples (99 unique stars) remain.

\subsection{ePSF Modelling}

\begin{figure}
    \centering
    \includegraphics[width=0.9\linewidth]{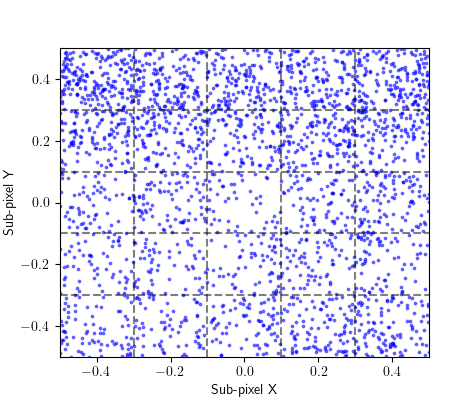}
    \caption{The sub-pixel distribution of the centroids of stars within the sample used to construct the ePSF model for the GJW-SA H-$\alpha$ camera. Black dashed lines show gridsection boundaries for an oversampling of five.}
    \label{fig:subpixel_samples}
\end{figure}

In this section, we construct the ePSF model using the filtered GJW-SA dataset described above. We first need to determine which oversampling parameter to use in the modelling.
Figure~\ref{fig:subpixel_samples} shows the sub-pixel distribution of the centre-of-mass centroid of each of the stars. With an oversampling of five, there are an average of 100 samples per gridsection and the minimum number of samples in a single gridsection is 46. This number of samples will be sufficient to accurately estimate the ePSF value at each of the gridpoints with the polynomial fitting method described in Section~\ref{subsec:gridpoint_estimation}. For our ePSF model, we use the RBF cubic interpolator from our investigations in Section~\ref{subsec:interpolation} implemented with SciPy's \texttt{RectBivariateSpline} class \citep{SciPy_2020}. No smoothing is applied to the ePSF model generated. We run the ePSF modelling routine through 10 iterations, with each iteration updating the measured stellar parameters and the ePSF model. 

\begin{figure*}
    \centering

    \begin{subfigure}[t]{0.45\linewidth}
        \centering
        \includegraphics[width=\linewidth]{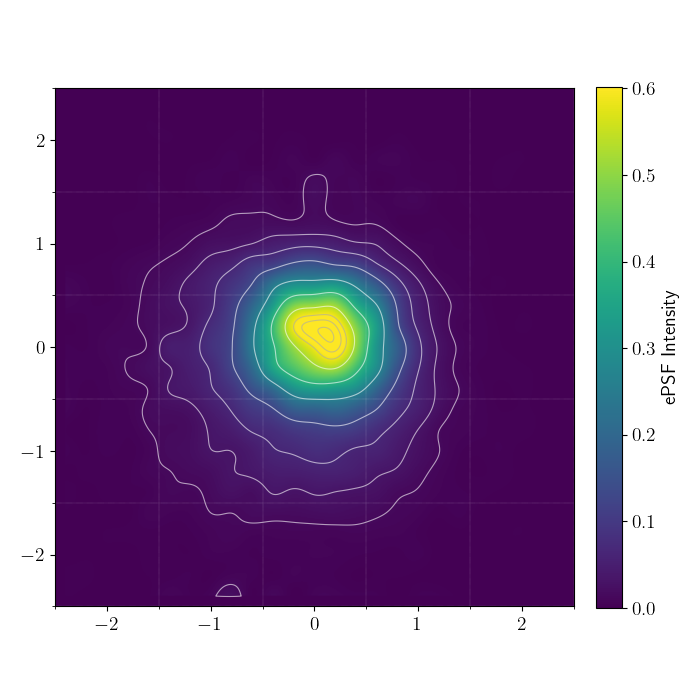}
        \caption{ePSF model (no smoothing).}
        \label{fig:epsf_model}
    \end{subfigure}
    \hfill
    \begin{subfigure}[t]{0.45\linewidth}
        \centering
        \includegraphics[width=\linewidth]{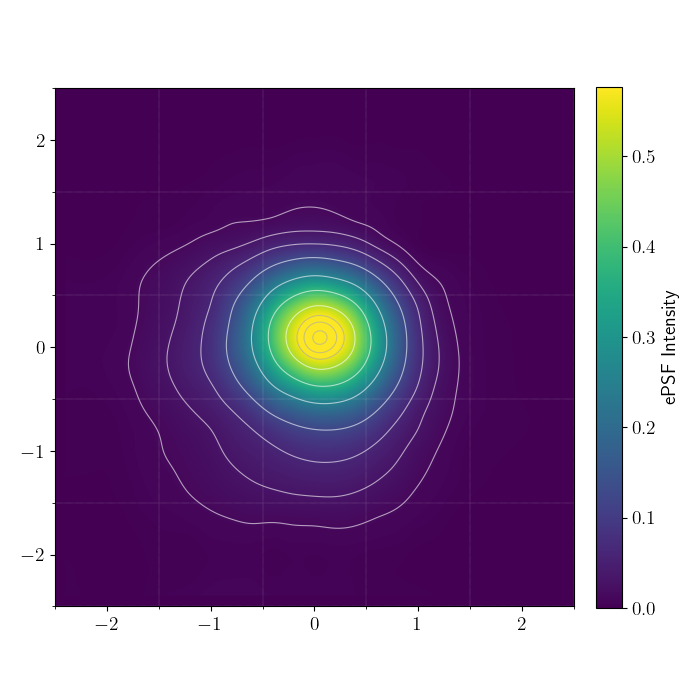}
        \caption{Smoothed ePSF model (5×5 cubic kernel).}
        \label{fig:epsf_model_smoothed}
    \end{subfigure}

    \vspace{1em}

    \begin{subfigure}[t]{0.45\linewidth}
        \centering
        \includegraphics[width=\linewidth]{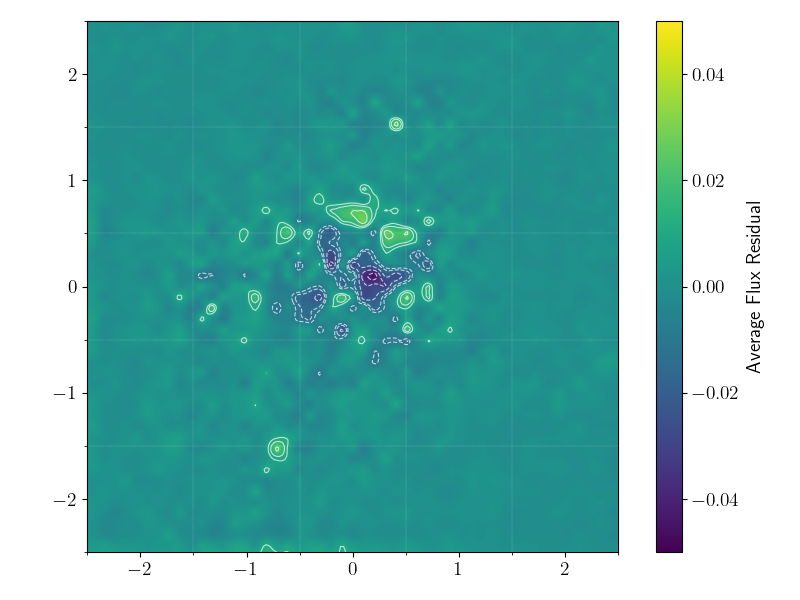}
        \caption{Residuals (unsmoothed model).}
        \label{fig:epsf_residuals}
    \end{subfigure}
    \hfill
    \begin{subfigure}[t]{0.45\linewidth}
        \centering
        \includegraphics[width=\linewidth]{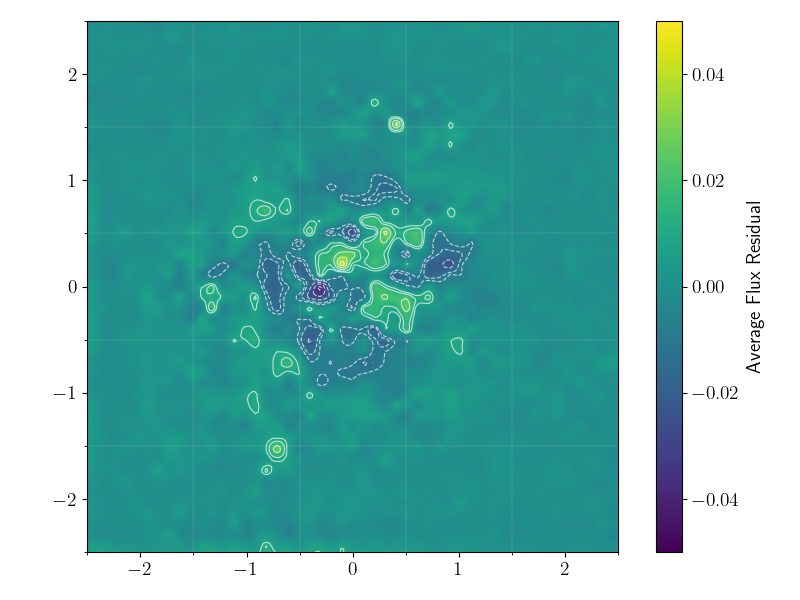}
        \caption{Residuals (smoothed model).}
        \label{fig:epsf_residuals_smoothed}
    \end{subfigure}

    \caption{Comparison of ePSF models generated using our updated modelling routine presented in this paper, and their corresponding residuals. \textbf{(a)} The ePSF model generated with an oversampling of 5 and no smoothing. \textbf{(b)} The same ePSF model with a 5×5 cubic smoothing kernel applied. \textbf{(c)} Residuals between the unsmoothed model and the set of sample star images. \textbf{(d)} Residuals between the smoothed ePSF model and the same dataset. In all panels, contour levels are shown in white and grey.}
    \label{fig:epsf_model_comparison}
\end{figure*}

Figure~\ref{fig:epsf_model} shows the resulting ePSF model. The residuals between this ePSF model and the stellar samples used to construct it are shown in Figure~\ref{fig:epsf_residuals}. These residuals demonstrate how accurately our ePSF model represents the star samples. In an ideal case, the residuals would consist only of small random noise. However, we observe some larger residuals---up to approximately 10\% of the ePSF peak intensity---and a systematic pattern around the peak of the ePSF model, suggesting that the model slightly overestimates the stellar flux in this region. 
These residuals likely originate from imperfections in the star samples themselves. The flux recorded by individual pixels in astronomical images contain multiple sources of error, and it is unlikely that all of these have been perfectly accounted for. This means the ePSF samples from stellar image pixels have non-negligible uncertainties, which limit the precision of the ePSF model and produce the residuals seen in Figure~\ref{fig:epsf_residuals}. 

We attempt to mitigate this issue by applying smoothing to the ePSF model. We generate the ePSF with the same configuration as previously, but with the addition of using a 5$\times$5 cubic smoothing kernel. Figure~\ref{fig:epsf_model_smoothed} shows the resulting ePSF model and Figure~\ref{fig:epsf_residuals_smoothed} show the residuals between this model and the star samples. The smoothing removes some of the fine scale structure in the ePSF model, and generally makes the model more circularly symmetric. There is no significant change in the scale of the residuals for this smoothed model, and there is not obvious evidence of over-smoothing, for which we would expect a very structured residual. 



\begin{figure}
    \centering
    \begin{subfigure}[t]{0.9\linewidth}
        \centering
        \includegraphics[width=\linewidth]{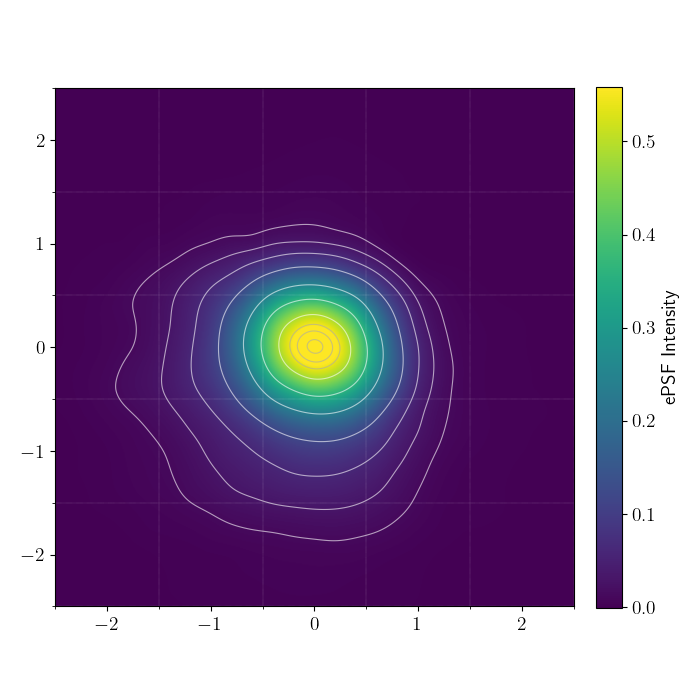}
        \caption{Smoothed ePSF model (\texttt{photutils}).}
        \label{fig:photutils_epsf_model}
    \end{subfigure}

    \vspace{1em}

    \begin{subfigure}[t]{\linewidth}
        \centering
        \includegraphics[width=\linewidth]{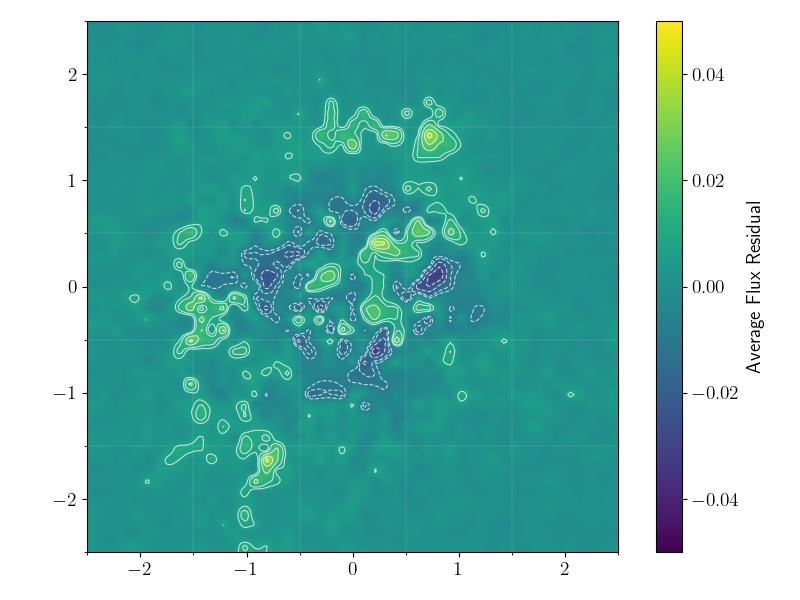}
        \caption{Residuals (smoothed \texttt{photutils} model).}
        \label{fig:photutils_epsf_residuals}
    \end{subfigure}

    \caption{\textbf{(a)} The ePSF model generated using \texttt{photutils.psf} with an oversampling of 5 and a 5×5 cubic smoothing kernel applied. \textbf{(b)} The corresponding residuals between this ePSF model and the set of star sample images used to generate it. In both panels, contour levels are shown in white and grey.}
    \label{fig:photutils_epsf_model_comparison}
\end{figure}

We compare this to the ePSF model generated from the same star samples but using the current implementation of the \texttt{photutils.psf} subpackage \citep{Bradley_2024}. As previously, the ePSF model is generated with an oversampling of five and with a 5$\times$5 cubic smoothing kernel. The default third degree bivariate spline interpolator is used for this model. We show the ePSF model after 10 iterations in Figure \ref{fig:photutils_epsf_model}, and its residuals with the star samples in Figure \ref{fig:photutils_epsf_residuals}. 
For this ePSF model, we observe residuals with similar scale to those in Figure~\ref{fig:epsf_residuals_smoothed}, however they are much tightly clustered around the peak of the ePSF. This means there are more significant residuals at parts of the ePSF model with relatively lower flux. This indicates that the ePSF modelling using the standard implementation of the code is less capable of accurately constraining our ePSF. We measure the widths of the two smoothed ePSF models at 90\% of their peak values along the $x=0$ and $y=0$ axes. We find that the \texttt{photutils} ePSF is 4\% wider along $x=0$ and 17\% wider along $y=0$ than our ePSF model in Figure~\ref{fig:epsf_model_smoothed}. This, combined with the shape of its residuals, indicates the \texttt{photutils} ePSF modelling may be overestimating the width of the ePSF.

\begin{figure*}
    \centering
    \begin{subfigure}[b]{0.48\textwidth}
        \centering
        \includegraphics[width=\textwidth]{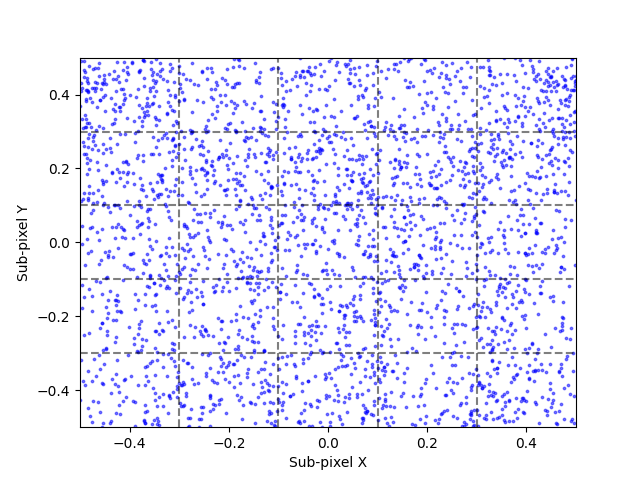}
        \caption{ePSF-fitted sub-pixel centroids (this work).}
        \label{fig:myepsf_subpixel_positions}
    \end{subfigure}
    \hfill
    \begin{subfigure}[b]{0.48\textwidth}
        \centering
        \includegraphics[width=\textwidth]{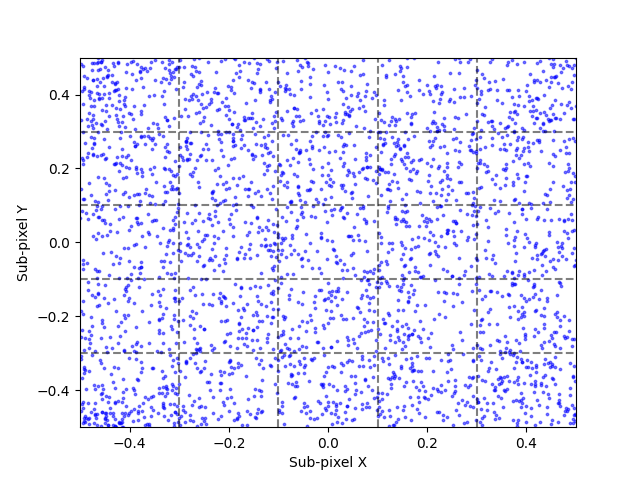}
        \caption{ePSF-fitted sub-pixel centroids (default \texttt{photutils}).}
        \label{fig:photutils_subpixel_positions}
    \end{subfigure}
    \caption{The sub-pixel distribution of the centroids of stars as measured with ePSF-fitting photometry. \textbf{(a)} Centroids measured with our improved ePSF model (Figure~\ref{fig:epsf_model_smoothed}). \textbf{(b)} Centroids measured with the ePSF model using the default \texttt{photutils.psf} subpackage (Figure~\ref{fig:photutils_epsf_model}). Black dashed lines show gridsection boundaries for an oversampling of five.}
    \label{fig:sub-pixel_dist_comparison}
\end{figure*}

We next evaluate the accuracy of ePSF-fitting photometry using the models generated above and compare it to aperture photometry to assess whether the ePSF modelling removes pixel-phase systematics. We use the same subset of 31 images that was used to construct the ePSF models and apply the ePSF fitting to all stars within the region and flux range for which the ePSF models were created. 
We first consider the pixel-phase errors on the measured centroid positions of the stars. In Figure~\ref{fig:subpixel_samples}, we see that the sub-pixel positions of stars measured with a centre-of-mass centroid are not uniformly distributed, indicating the presence of pixel-phase errors. In Figure~\ref{fig:sub-pixel_dist_comparison} we show the distribution of the sub-pixel centroids of the same stars as measured with ePSF-fitting, using the improved ePSF from this work and the ePSF constructed with the default \texttt{photutils.psf} subpackage. For both cases, we see that the sub-pixel centroids are much more uniformly distributed than the centre-of-mass centroid case, showing that the pixel-phase errors in the position measurements are reduced. There remain some non-uniformities in these distributions, for example a clustering in the bottom left corner in Figure~\ref{fig:photutils_subpixel_positions}. This indicates some remaining errors in the ePSF models, which result in small pixel-phase systematics in the measured positions of stars.

\begin{figure*}
    \centering
    \includegraphics[width=\textwidth]{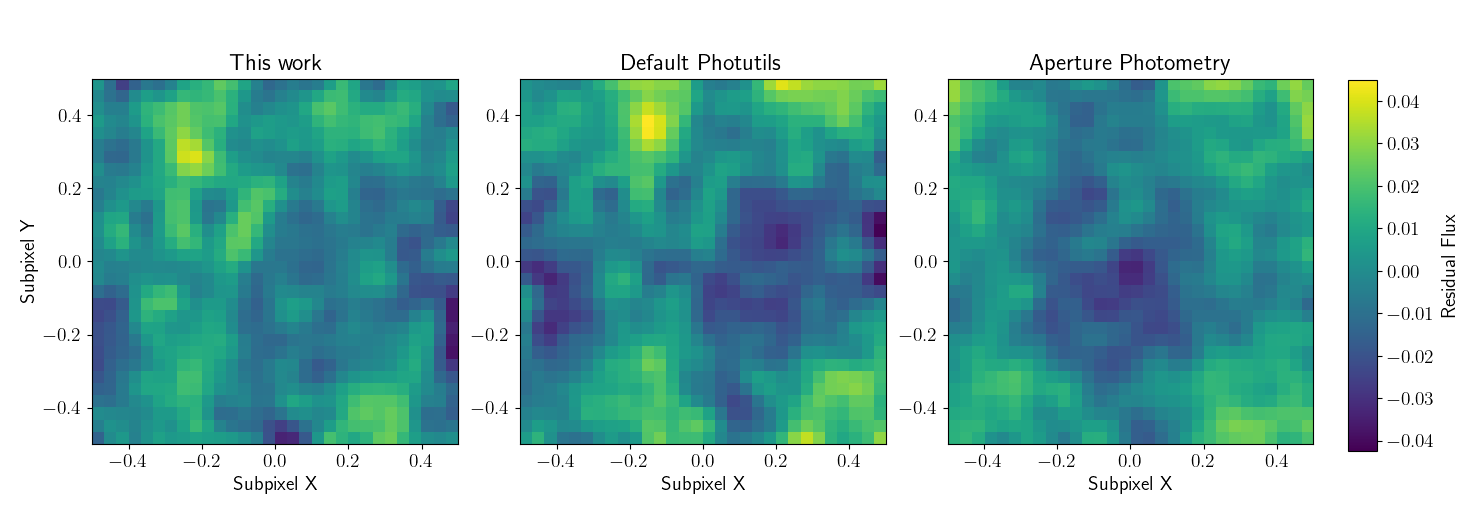}
    \caption{Comparison of flux pixel-phase error (PPE) plots for ePSF-fitting photometry using our improved ePSF model, the default \texttt{photutils.psf} model, and aperture photometry. The plots show 2D maps of the average flux residual with respect to sub-pixel centroid position of a star.}
    \label{fig:ppe_comparisons}
\end{figure*}

To visualise the pixel-phase dependencies in the measured fluxes of stars, we produce two-dimensional maps of the mean flux residuals as a function of sub-pixel centroid position. Systematic patterns in these maps indicate residual inaccuracies in the ePSF model.
Figure~\ref{fig:ppe_comparisons} compares the flux pixel-phase errors from our ePSF model, the default \texttt{photutils.psf} implementation, and aperture photometry. The pixel-phase maps in this figure have been slightly smoothed to highlight evidence of systematic trends. For the aperture photometry fluxes, we observe that the flux residuals are more negative towards the centre of a pixel and more positive close to the corners of a pixel. This indicates the presence of some non-uniform intra-pixel sensitivity profile for the pixels of this camera. We see similar scale residuals in the pixel-phase maps for our ePSF model, however we no longer see the same pattern in the residuals. There is still some structure in the pixel-phase residual map which is due to the inaccuracies remaining in the ePSF model. The flux residuals are strongest for the \texttt{photutils} ePSF, and show large patches of clear positive and negative systematics. This indicates that this ePSF model has some large deviations from the true ePSF, which cause systematic errors in recorded fluxes which can be even worse than those for aperture photometry.

The ePSF modelling for the GJW commensal cameras is a particularly challenging application, due to the variations in the instrumental PSF (position dependence, brighter-fatter effect, atmospheric seeing fluctuations) and many sources of error on the recorded pixel values of stars within the images. Although we have not been able to generate a perfect ePSF model in this work, our investigations into the ePSF modelling routine and updates to the Python implementation have enabled us to improve on the accuracy of the ePSF model compared with the original modelling routine in the \texttt{photutils} package. Work into modelling the ePSF for each of the GJW commensal cameras, accounting for instrumental PSF variations, is ongoing and will be presented in Godden et~al. (in prep).

\section{Conclusions}

Accurate modelling of the effective point spread function (ePSF) is essential for achieving reliable photometric and astrometric precision, especially in undersampled imaging regimes. In this work, we have conducted a systematic investigation into the accuracy of ePSF modelling, following the method of \citet{Anderson_King_2000} and its implementation in the open-source \texttt{photutils.psf} subpackage of \texttt{Astropy} \citep{Bradley_2024}. Using synthetic ePSFs to generate simulated datasets, we quantified how each component of the ePSF-building routine---oversampling, interpolation, gridpoint estimation, smoothing, star sample distribution, degeneracy breaking, and dither patterns---affect the accuracy of the ePSF model and resulting photometric measurements.

Through controlled experiments, we identified several practical strategies for optimising ePSF construction. 
\begin{enumerate}
  \item \textbf{Oversampling.} The oversampling factor strongly affects model and photometric accuracy. Use the highest oversampling permitted by the available star samples; for near-Gaussian ePSFs we recommend at least $\sim4$ gridpoints per FWHM.
  \item \textbf{Interpolation.} A radial-basis-function interpolator with a cubic kernel (RBF cubic) gives the most accurate interpolation between gridpoints, especially for undersampled cases.
  \item \textbf{Gridpoint estimation.} Two-dimensional polynomial surface fits outperform sigma-clipped mean/median estimators, particularly when ePSF samples are sparse or unevenly distributed.
  \item \textbf{Smoothing.} Apply smoothing cautiously to suppress noise in empirical gridpoint values; over-smoothing flattens real structure and induces pixel-phase biases.
  \item \textbf{Sample size.} Monte Carlo tests quantify how many randomly distributed star samples are required to accurately estimate the gridpoints of the ePSF model; in practice, always inspect the sub-pixel distribution of your samples before fixing the oversampling.
  \item \textbf{Dithering.} For dedicated dithered observations, uniform sampling of the sub-pixel space (ideally matching the oversampling grid) is far more effective than random dithers.
\end{enumerate}

Importantly, we demonstrate that degeneracy breaking must constrain \emph{both} positions \emph{and} fluxes of stars across dithered frames. Constraining positions alone, the current default in \texttt{photutils.psf}, does not remove the flux–position degeneracy introduced by intra-pixel sensitivity. Constraining both the fluxes and positions is necessary to prevent intra-pixel sensitivity variations from introducing systematic errors in the ePSF model and in subsequent photometric and astrometric measurements. This insight provides a clear direction for improving future implementations of \texttt{photutils.psf} and similar ePSF frameworks.

We further applied our refined ePSF-modelling framework to real observational data from the Global Jet Watch (GJW) commensal cameras. Despite the challenges of undersampled and noisy images, the ePSF models constructed with our modified routine produced photometry with reduced pixel-phase systematics compared to the default \texttt{photutils.psf} implementation. While residual scatter remains due to the highly undersampled nature of the data, these results demonstrate the feasibility and benefits of accurate ePSF modelling for real ground-based observations.

Overall, this work provides both theoretical and practical guidance for optimising ePSF modelling routines. The simulation framework, code improvements, and recommendations presented here form a foundation for robust ePSF modelling across diverse instruments. 
This is a step towards understanding how the ePSF modelling could be fully automated so that minimal human interaction is required to generate an ePSF model for a new instrument. 

\section*{Acknowledgements}

The authors acknowledge support from the Science and Technology Facilities Council (STFC), and from St Cross College, University of Oxford, through the St Cross Astrophysics scholarship. This research made use of Astropy, a community-developed core Python package for Astronomy \citep{astropy:2013, astropy:2018, astropy:2022}, and SciPy, an open-source Python library for scientific computing \citep{SciPy_2020}.

\section*{Data Availability}

The code used to generate our simulated data and the code used to run the experiments in this paper can be provided upon reasonable request to the authors. Our modified version of the \texttt{photutils.psf} subpackage of Astropy will be submitted for inclusion in the official open-source release.
 



\bibliographystyle{mnras}
\bibliography{example} 




\appendix

\section{Synthetic ePSF Derivations}
\label{ap:analytic_epsfs}

In our simulations, we use a \textit{flat with gaps} intra-pixel sensitivity profile. This corresponds to a two-dimensional top-hat function of half-width $w$ (so the sensitive side-length is $l=2w$)
\begin{equation}
    R(x, y) =
    \begin{cases}
      1, & \text{if } |x| \le w \text{ and } |y| \le w, \\
      0, & \text{otherwise.}
    \end{cases}
    \label{eqn:pixel_response_function}
\end{equation}
The half-width $w$ relates to the fill factor (the ratio of a pixel’s light-sensitive area to its total area) as $\mathrm{FF} = 4w^2$.  
A \textit{flat and flush} pixel model (no gaps, $\mathrm{FF} = 1$) corresponds to $w = 0.5$, whereas a 9:1 ratio between collecting-area width and inter-pixel gap corresponds to $w = 0.45$ ($\mathrm{FF} = 0.81$).  

Below we give derivations of the ePSF, $\Psi_{\mathrm{E}}(i,j)$, for instrumental PSF models of interest using this intra-pixel sensitivity profile. These provide the basis for the simulated ePSFs used throughout this work.

\subsection{Circular Gaussian PSF}

We begin with a circular Gaussian PSF with standard deviation~$\sigma$:
\begin{equation}
    \Psi_{\mathrm{I}}(x,y) = \frac{1}{2\pi\sigma^2}
    \exp\!\left[-\frac{x^2 + y^2}{2\sigma^2}\right] .
\end{equation}

Following Equation~\ref{eqn:epsf}, the ePSF can be written as:
\begin{equation}
    \Psi_{\mathrm{E}}(i, j) = \frac{1}{2\pi\sigma^2}
    \int_{i-w}^{i+w} \int_{j-w}^{j+w}
    \exp\!\left[-\frac{(x - x_*)^2 + (y - y_*)^2}{2\sigma^2}\right]
    \, \mathrm{d}x\,\mathrm{d}y ,
\end{equation}
where $(i,j)$ are pixel coordinates and $(x_*, y_*)$ is the centroid of the star.  
Because the Gaussian is separable in $x$ and $y$, the ePSF integral factorises and can be evaluated in closed form using the error function,
\begin{equation}
    \operatorname{erf}(z) = \frac{2}{\sqrt{\pi}} \int_{0}^{z} e^{-t^2}\, \mathrm{d}t .
\end{equation}
We obtain the ePSF
\begin{multline}
    \Psi_{\mathrm{E}}(i, j)
    = \frac{1}{4}
    \bigg[
      \operatorname{erf}\!\left(\frac{i + w - x_*}{\sqrt{2}\,\sigma}\right)
      - \operatorname{erf}\!\left(\frac{i - w - x_*}{\sqrt{2}\,\sigma}\right)
    \bigg] \\
    \times
    \bigg[
      \operatorname{erf}\!\left(\frac{j + w - y_*}{\sqrt{2}\,\sigma}\right)
      - \operatorname{erf}\!\left(\frac{j - w - y_*}{\sqrt{2}\,\sigma}\right)
    \bigg] .
    \label{eqn:circular_gaussian_epsf}
\end{multline}

In our simulations, the error function terms are evaluated using the \texttt{scipy.special.erf} implementation.

\subsection{Elliptical Gaussian PSF}

The same approach can be applied to an elliptical Gaussian PSF with different widths $\sigma_x$ and $\sigma_y$:
\begin{multline}
    \Psi_{\mathrm{E}}(i, j)
    = \frac{1}{4}
    \bigg[
      \operatorname{erf}\!\left(\frac{i + w - x_*}{\sqrt{2}\,\sigma_x}\right)
      - \operatorname{erf}\!\left(\frac{i - w - x_*}{\sqrt{2}\,\sigma_x}\right)
    \bigg] \\
    \times
    \bigg[
      \operatorname{erf}\!\left(\frac{j + w - y_*}{\sqrt{2}\,\sigma_y}\right)
      - \operatorname{erf}\!\left(\frac{j - w - y_*}{\sqrt{2}\,\sigma_y}\right)
    \bigg] .
    \label{eqn:asymmetric_gaussian_epsf}
\end{multline}

\subsection{Multi-Component Gaussian PSF}

For an instrumental PSF represented as a superposition of $k$ Gaussian components with relative amplitudes $a_k$, the ePSF becomes:
\begin{equation}
    \Psi_{\mathrm{E}}(i, j)
    = \frac{1}{4}
      \frac{\sum_k g_k(i, j)}{\sum_k a_k} ,
\end{equation}
where each component $g_k$ is given by:
\begin{multline}
    g_k(i, j) = a_k
    \bigg[
      \operatorname{erf}\!\left(\frac{i + w - (x_* + x_k)}{\sqrt{2}\,\sigma_{k,x}}\right)
      - \operatorname{erf}\!\left(\frac{i - w - (x_* + x_k)}{\sqrt{2}\,\sigma_{k,x}}\right)
    \bigg] \\
    \times
    \bigg[
      \operatorname{erf}\!\left(\frac{j + w - (y_* + y_k)}{\sqrt{2}\,\sigma_{k,y}}\right)
      - \operatorname{erf}\!\left(\frac{j - w - (y_* + y_k)}{\sqrt{2}\,\sigma_{k,y}}\right)
    \bigg] .
\end{multline}

\subsection{Rotated Elliptical Gaussian PSF}
\label{subsec:rotated_gaussian}

To represent a PSF whose principal axes are not aligned with the detector pixel grid, we use a rotated elliptical Gaussian model,
\[
\Psi_{\mathrm{R}}(x,y)
= \frac{1}{2\pi\sqrt{|\Sigma|}}
\exp\!\Big[-\tfrac{1}{2}(\mathbf{r}-\boldsymbol{\mu})^{T}\Sigma^{-1}(\mathbf{r}-\boldsymbol{\mu})\Big],
\]
where $\mathbf{r}=(x,y)^{T}$ and $\boldsymbol{\mu}=(x_*,y_*)^{T}$ are the position and centroid vectors, and $\Sigma$ is the $2\times2$ covariance matrix describing the PSF shape and orientation.  
The covariance matrix can be written as
\[
\Sigma = R(\theta)
\begin{pmatrix}
\sigma_1^2 & 0\\[4pt]
0 & \sigma_2^2
\end{pmatrix}
R(\theta)^{T},
\]
where $\sigma_1$ and $\sigma_2$ are the standard deviations along the PSF's principal axes, and $R(\theta)$ is the rotation matrix
\[
R(\theta) =
\begin{pmatrix}
\cos\theta & -\sin\theta\\[4pt]
\sin\theta & \phantom{-}\cos\theta
\end{pmatrix}.
\]
This formulation reduces to the axis-aligned elliptical Gaussian when $\theta = 0$ and to the circular Gaussian when $\sigma_1 = \sigma_2$.

The ePSF is obtained by integrating the bivariate Gaussian over the rectangular pixel area. This integral can be expressed as a combination of bivariate normal cumulative distribution functions (CDFs) evaluated at the pixel boundaries:
\[
\Psi_{\mathrm{E,R}}(i,j)
= F(i+w,j+w) - F(i-w,j+w)
- F(i+w,j-w) + F(i-w,j-w),
\]
where $F(x,y)$ denotes the bivariate normal CDF for mean~$\boldsymbol{\mu}$ and covariance~$\Sigma$,
\[
F(x,y) =
\frac{1}{2\pi\sqrt{|\Sigma|}}
\int_{-\infty}^{x}\!\int_{-\infty}^{y}
\exp\!\Big[-\tfrac{1}{2}(\mathbf{r}'-\boldsymbol{\mu})^{T}\Sigma^{-1}(\mathbf{r}'-\boldsymbol{\mu})\Big]
\,\mathrm{d}y'\,\mathrm{d}x'.
\]

In practice, the bivariate normal CDF can be evaluated efficiently using standard numerical routines (e.g.\ \texttt{scipy.stats.multivariate\_normal.cdf}). 

\subsection{Skewed Gaussian}

To represent a PSF with a tail preferentially on one side we use a skew-normal–type model that multiplies a two-dimensional Gaussian by a one-dimensional skew factor in \(x\). We can represent this skew factor using the standard normal CDF
\[
\Phi(z) = \frac{1}{\sqrt{2\pi}} \int_{-\infty}^{z} e^{-t^{2}/2}\, \mathrm{d}t .
\]
The skewed Gaussian PSF can then be written as
\begin{equation}
\Psi_{\mathrm{skew}}(x,y)
= \frac{1}{2\pi\sigma_x\sigma_y}
\exp\!\left[-\frac{(x-x_*)^2}{2\sigma_x^2}-\frac{(y-y_*)^2}{2\sigma_y^2}\right]
\Phi\!\Big(\frac{\eta\,x}{\sigma_x}\Big),
\label{eq:skew_model}
\end{equation}
where \(\eta\) is the skew parameter (\(\eta>0\) produces a tail on the \(+x\) side).
Sine the skew factor depends only on \(x\), the ePSF integral factorises:
\[
\Psi_{\mathrm{E}}(i,j) \;=\; I_x(i)\,I_y(j).
\]
The \(y\)-part can be written analytically as
\begin{align}
I_y(j) &= \int_{j-w}^{j+w}\frac{1}{\sqrt{2\pi}\sigma_y}
\exp\!\Big[-\frac{(y-y_*)^2}{2\sigma_y^2}\Big]\,\mathrm{d}y
\label{eq:Iy_exact}\\
&= \frac{1}{2}\Bigg[
\operatorname{erf}\!\Big(\frac{j+w-y_*}{\sqrt{2}\sigma_y}\Big)
-\operatorname{erf}\!\Big(\frac{j-w-y_*}{\sqrt{2}\sigma_y}\Big)
\Bigg]. \nonumber
\end{align}
This leaves the \(x\)-part as a one-dimensional integral
\begin{equation}
I_x(i) \;=\; \int_{i-w}^{i+w}\frac{1}{\sqrt{2\pi}\sigma_x}
\exp\!\Big[-\frac{(x-x_*)^2}{2\sigma_x^2}\Big]\,
\Phi\!\Big(\eta\,\frac{x-x_0}{\sigma_x}\Big)\,\mathrm{d}x .
\label{eq:Ix_integral}
\end{equation}

For our simulations, we evaluate \(I_x(i)\) with a robust one-dimensional integrator  (e.g.\ \texttt{scipy.integrate.quad}). 

\section{Modifications to \texttt{photutils.psf} subpackage}
\label{ap:code_changes}

This appendix describes the modifications and extensions made to the \texttt{photutils.psf} subpackage in order to align the ePSF modelling routine implemented in the \texttt{EPSFBuilder} class more closely with that described by \citet{Anderson_King_2000}, and to enable additional flexibility in the ePSF modelling and fitting processes.

\subsection{Core Structural Changes}

\paragraph*{Replacement of Legacy ePSF Model.}
Both the \texttt{EPSFFitter} and \texttt{EPSFBuilder} classes currently implement the \texttt{\_LegacyEPSFModel} class to represent an ePSF model. This class is a copy of the deprecated \texttt{EPSFModel} class, which has since been superseded by the \texttt{ImagePSF} class. In our modifications, the \texttt{EPSFBuilder} and \texttt{EPSFFitter} classes were refactored to use the \texttt{ImagePSF} class, or its subclasses, to represent the ePSF model. During this process, we identified an error in the \texttt{\_create\_initial\_epsf} method of \texttt{EPSFBuilder}, in which the ePSF centre was incorrectly calculated due to a misunderstanding of Python’s zero-based indexing. This error caused numerical instabilities in the ePSF modelling routine once the \texttt{ImagePSF} implementation was adopted. After correcting the centre calculation, the routine produced stable, convergent ePSF models.

\paragraph*{Introduction of Interpolator Subclasses.}
We define a set of subclasses of the \texttt{ImagePSF} class that implement alternative interpolation methods for evaluating the ePSF at arbitrary, including sub-pixel, positions. In addition to the default third degree \texttt{RectBivariateSpline} interpolator used in \texttt{ImagePSF}, we provide subclasses that use \texttt{SmoothBivariateSpline} for smoothed spline interpolation, \texttt{RectBivariateSpline} with user-defined polynomial degree and smoothing, and \texttt{RBFInterpolator} for radial basis function interpolation with a user-defined kernel (i.e. `cubic'). 

\paragraph*{Introduction of New Fitter Class.}
The \texttt{EPSFFitter} class accepts a \texttt{fitter} parameter to specify which \texttt{astropy.modeling.fitting.Fitter} object to use when fitting the ePSF model to a star image. The default is the \texttt{TRFLSQFitter} (Trust Region Reflective Least Squares) from \texttt{scipy.optimize.least\_squares}. However, we found that this fitter can produce ill-conditioned covariance matrices when fitting parameters such as flux and centroid position that differ by several orders of magnitude, leading to numerical instability. If we fit for the natural logarithm of the flux rather than the linear flux, giving more stable and consistent parameter uncertainties without introducing bias in the recovered flux. This adjustment is particularly important for undersampled images, where only a few pixels constrain each stellar profile and the fit is especially sensitive to numerical issues.

\begin{figure*}
  \centering

  \begin{subfigure}[t]{0.45\linewidth}
    \centering
    \includegraphics[width=\linewidth]{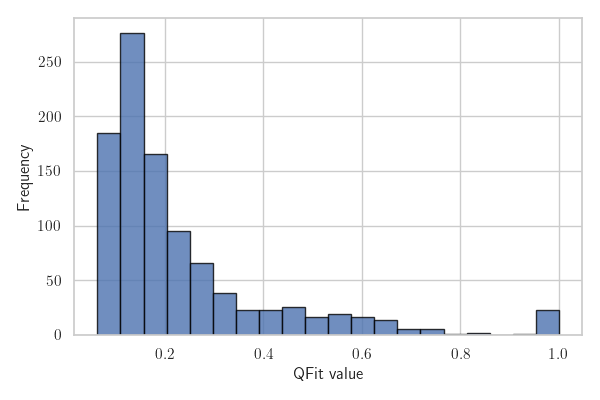}
    \caption{Histogram of QFit results using \texttt{TRFLSQFitter}}
    \label{fig:histA}
  \end{subfigure}
  \hfill
  \begin{subfigure}[t]{0.45\linewidth}
    \centering
    \includegraphics[width=\linewidth]{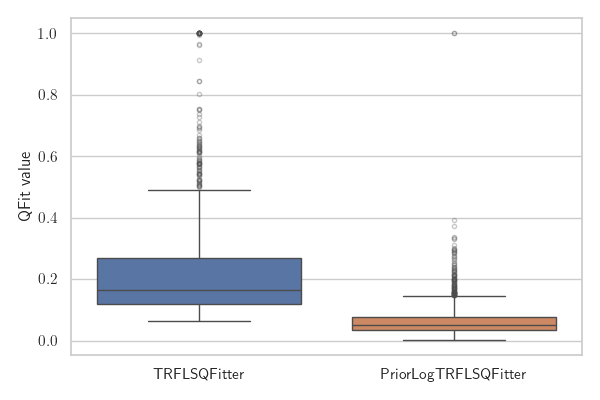}
    \caption{Boxplots of QFit results comparing \texttt{TRFLSQFitter} and \texttt{PriorLogTRFLSQFitter}. Grey circles show outliers.}
    \label{fig:boxAB}
  \end{subfigure}

  \vspace{1em}

  \begin{subfigure}[t]{0.45\linewidth}
    \centering
    \includegraphics[width=\linewidth]{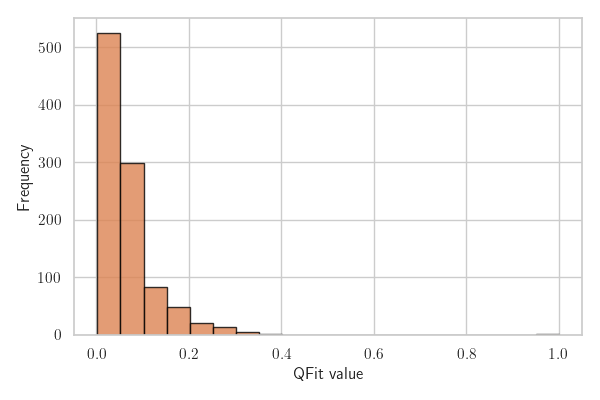}
    \caption{Histogram of QFit results using \texttt{PriorLogTRFLSQFitter}}
    \label{fig:histB}
  \end{subfigure}
  \hfill
  \begin{subfigure}[t]{0.45\linewidth}
    \centering
    \includegraphics[width=\linewidth]{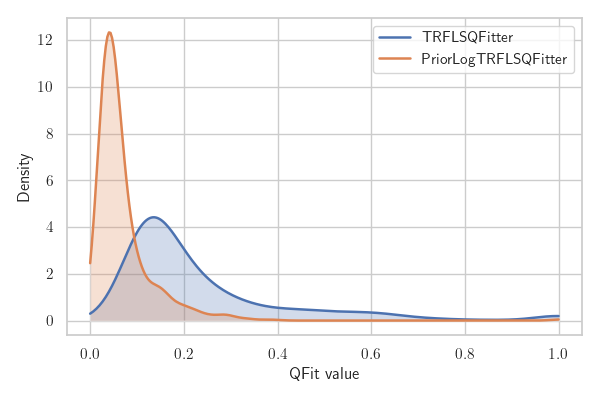}
    \caption{KDE / density comparison of QFit results using \texttt{TRFLSQFitter} and \texttt{PriorLogTRFLSQFitter}.}
    \label{fig:kdeAB}
  \end{subfigure}

  \caption{Comparison of QFit distributions from fitting of 1000 simulated stars with their known ePSF using different fitting methods.}
  \label{fig:qfit_comparison}
\end{figure*}

We define a new TRF-like fitter class, \texttt{PriorLogTRFLSQFitter}, which fits the $\log(\mathrm{flux})$ internally, supports Gaussian priors, and returns covariances in both internal $(\log(\mathrm{flux}), x_0, y_0)$ and physical $(\mathrm{flux}, x_0, y_0)$ spaces. Figures~\ref{fig:histA}–\ref{fig:kdeAB} show validation tests comparing this fitter with the standard \texttt{TRFLSQFitter}. Using simulated stars generated from ePSF Model~A (Figure~\ref{fig:input_epsfs}) with fluxes randomly drawn between 10,000 and 100,000 counts, we fit each star with both methods and evaluate the results using the QFit statistic (the sum of absolute residuals between the data and fitted model). The \texttt{PriorLogTRFLSQFitter} consistently produces smaller QFit values, indicating more accurate and robust fits.

\paragraph*{Introduction of Synthetic ePSF Models.}

The synthetic ePSF models described in Appendix~\ref{ap:analytic_epsfs} are implemented as subclasses of the \texttt{Fittable2DModel} class of the \texttt{astropy.modelling} subpackage, following the convention of the `PRF' models defined in the \texttt{photutils.psf} subpackage. The term `PRF' is equivalent to `ePSF' in this usage. These new classes make it possible to simulate ePSFs with a \textit{flat with gaps} intra-pixel sensitivity profile, making it possible to perform investigations into the effect of non-uniform pixel response, as we have done in Sections~\ref{subsec:degeneracy_breaking} and \ref{subsec:dither_pattern}.

\subsection{Modifications to ePSF Modelling Routine}

We have introduced several modifications to the \texttt{EPSFBuilder} class to provide greater flexibility in configuring the ePSF modelling routine. The original behaviour and settings remain accessible by overriding the default parameters during class initialisation.

\paragraph*{Alternative Gridpoint Estimation Methods.}

The gridpoint estimation methods tested in Section~\ref{subsec:gridpoint_estimation} are implemented in the \texttt{EPSFBuilder} class. The desired method can be specified with the \texttt{gridpoint\_estimation} parameter on class initialisation. 


\paragraph*{Implementation of ePSF Normalisation.}

We implement normalisation of the ePSF model on each iteration in the modelling routine. The ePSF is normalised such that a star with a flux of 1 centred at the middle of a pixel will have an aperture sum of its pixel values of 1. This normalisation can be skipped by setting the parameter \texttt{normalise\_epsf = False} on the \texttt{EPSFBuilder} class initialisation.

\paragraph*{Constraining Fluxes and Positions for Degeneracy Breaking.}

We define a new method of the \texttt{LinkedEPSFStars} class, \texttt{constrain\_fluxes}, to constrain the fluxes of linked stars to be their average value. This method is analogous to that for the stellar positions, \texttt{contrain\_centers}. We remove the call of the \texttt{constrain\_centers} method in the initialisation of the \texttt{LinkedEPSFStars} class, since there may be occasions where we don't want to automatically constrain parameters of linked stars. We instead constrain both the fluxes and positions of linked stars from within the \texttt{EPSFBuilder} class during the modelling of the ePSF.

\section{GJW Image Pre-Processing}
\label{ap:gjw_preprocessing}

\begin{figure*}
    \centering
    \includegraphics[width=\linewidth]{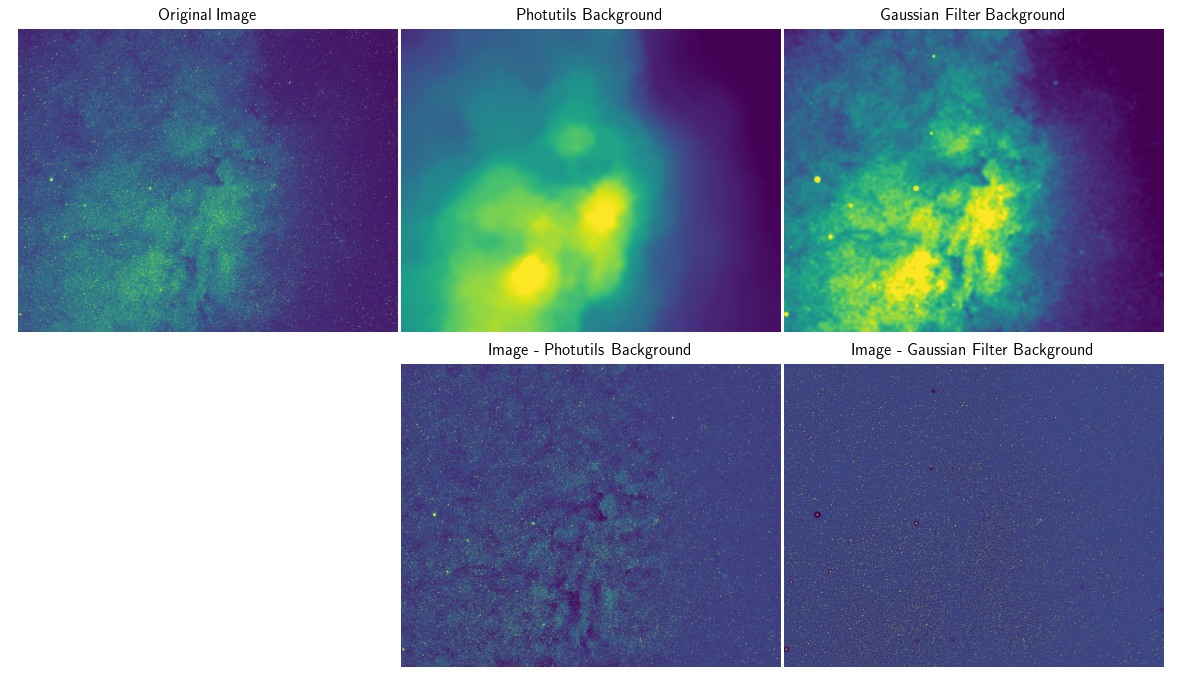}
    \caption{Comparison of background estimation methods using a single image of the V6595 Sgr field with the GJW-SA H-$\alpha$ camera. Left: Raw image. Middle: Background and subtracted image with \texttt{photutils} \texttt{Background2D} estimation. Right: Background and subtracted image with Gaussian filtering estimation.}
    \label{fig:backgrounds}
\end{figure*}

Prior to images being used in ePSF modelling, we perform some image processing steps:

\paragraph*{Instrumental Calibration.}

The raw images are instrumentally calibrated using bias, dark, and flat field images. For each of these calibrations, we observe a set of calibration images and produce an averaged \textit{master} calibration image. The raw images are then calibrated using the standard approach:

\begin{equation}
I_{\text{cal}} = \frac{I_{\text{raw}} - B - D}{F},
\end{equation}

where $B$ is the master bias image, $D$ is the master dark image (bias-subtracted and calibrated for exposure time), and $F$ is the normalised master flat field image (bias and dark calibrated).

\paragraph*{Background Subtraction.}

The raw images in this work are taken with an H-$\alpha$ filter, so we need to take care to remove the background due to diffuse nebular light which appears in the images. We find that the background estimation method in the \texttt{Background2D} class of the \texttt{photutils.background} Python subpackage does not accurately model the background and as a result some structured background light remains in the images. Instead, we decide to estimate the background using Gaussian filtering. 

Figure \ref{fig:backgrounds} shows a comparison between the background modelling using this method and with the \texttt{photutils.background.Background2D} method. In our background estimation, the large-scale background was estimated via a Gaussian convolution whose width was chosen to be large enough to smooth out flux from stars and high-frequency noise, yet small enough to preserve the smaller-scale variations of the nebulous background light. Masks were placed over each star to ensure that stellar flux did not contribute to the background estimate. A PSF-scale smoothed version of the background-subtracted image was then subtracted to obtain a high-frequency residual, from which the local RMS map was derived. To capture the smaller-scale structure in the background, we used a Gaussian smoothing kernel with a width of $\sigma \approx 6.8~\mathrm{px}$ (corresponding to $\mathrm{FWHM} \approx 16~\mathrm{px}$). This relatively narrow filter causes a small amount of flux from bright stars that are not fully masked to be included in the background model. This could be mitigated by identifying these bright stars and applying larger mask radii. However, since the ePSF modelling only uses stars within a specific flux range, the brightest stars are excluded from that analysis, and therefore the inclusion of a small fraction of their flux in the background model has no impact on our results in this work.

\paragraph*{Astrometric Calibration.}

After background calibration, we then calculate the World Coordinate System (WCS) transformation of the images using the Astrometry.net software \citep{Lang_2010}. This is essential for being able to crossmatch detections of the same stars across different frames.

\paragraph*{Source Detection.}

We use the \texttt{photutils detect\_sources} function in order to identify stars above a specified threshold in our images \citep{Bradley_2024}. Our threshold for detection is set such that there must be four connected pixels at a value of five standard deviations above the local background value.


\bsp	
\label{lastpage}
\end{document}